\documentclass[acmtog,nonacm]{acmart}

\usepackage{algorithmic}
\usepackage{enumitem}
\usepackage{booktabs} % For formal tables
\usepackage{bbding}
\usepackage{pifont} %?
\usepackage{wasysym} %?
\usepackage{utfsym} %?
\usepackage{fontawesome} %?

\usepackage[percent]{overpic} %?
\usepackage{color}
\usepackage{placeins} %?
\usepackage{makecell} %?
\usepackage{multirow}
\usepackage{amsmath}
\usepackage{amssymb}
\usepackage{color}
\usepackage{graphicx}
\usepackage[ruled]{algorithm2e} % For algorithms

\newcommand{\SQ}[1]{{\color{red}#1}}

\newcommand{\PF}[1]{{\color{purple}PF: #1}}

\setcopyright{acmcopyright}
\acmJournal{TOG}
\acmYear{2022}
\acmVolume{41}
\acmNumber{6}
\acmArticle{185}
\acmMonth{12}
\acmDOI{10.1145/3550454.3555453}

\begin{document}

% Title. 
% If your title is long, consider \title[short title]{full title} - "short title" will be used for running heads.
\title{SurfaceVoronoi: Efficiently Computing Voronoi Diagrams Over Mesh Surfaces with Arbitrary Distance Solvers}
%\subtitle{Retrieved from \url{https://www.siggraph.org/learn/instructions-authors} on June 5th, 2019}

% Authors.
\author{Shiqing Xin}
\affiliation{%
 \institution{Shandong University}
 \country{China}
 }
\email{xinshiqing@sdu.edu.cn}
\author{Pengfei Wang}
\affiliation{%
 \institution{Shandong University}
 \country{China}}
 \email{pengfei1998@foxmail.com}
 \author{Rui Xu}
\affiliation{%
 \institution{Shandong University}
 \country{China}}
 \email{xrvitd@163.com}
 \author{Dongming Yan}
\affiliation{%
% \institution{University of Chinese Academy of Sciences}
\institution{National Laboratory of Pattern Recognition (NLPR), Institute of Automation, Chinese Academy of Sciences, and School of AI, University of Chinese Academy of Sciences}
 \country{China}}
\email{yandongming@gmail.com}
 \author{Shuangmin Chen}
 \authornote{Shuangmin Chen is the corresponding author.}
\affiliation{%
 \institution{Qingdao University of Science and Technology}
 \country{China}}
 \email{csmqq@163.com}
 \author{Wenping Wang}
\affiliation{%
 \institution{Texas A\&M University}
 \country{USA}}
 \email{wenping@tamu.edu}
 \author{Caiming Zhang}
\affiliation{%
 \institution{Shandong University}
 \country{China}}
 \email{czhang@sdu.edu.cn}
 \author{Changhe Tu}
\affiliation{%
 \institution{Shandong University}
 \country{China}}
 \email{chtu@sdu.edu.cn}
 
% \makeatletter
% \let\@authorsaddresses\@empty
% \makeatother

% \author{Brittany Rowland-Smith}
% \affiliation{%
%   \institution{St. Olaf College}}
% \email{br-s@gmail.com}

% \author{Nicholas Badeeri}
% \affiliation{%
%   \institution{MathWorks, Inc.}
%   \email{badeeri@mathworks.com}
%   }

% \author{Andrew Joseph Foyt}
% \affiliation{%
%   \department{College of Engineering}
%   \institution{University of Houston}}
% \email{foyt_aj@uh.edu}

% This command defines the author string for running heads.
% \renewcommand{\shortauthors}{DeJohnette, Rowland-Smith, Badeeri, and Foyt}
%\renewcommand{\shortauthors}{Tagliasacchi et al.}

% abstract
\begin{abstract}
In this paper, we propose to compute Voronoi diagrams over mesh surfaces
driven by an arbitrary geodesic distance solver, assuming that the input is a triangle mesh as well as a collection of sites $\mathbf{P}=\{p_i\}_{i=1}^m$ on the surface. 
We propose two key techniques to solve this problem. 
First, as the partition is determined by minimizing the $m$ distance fields, each of which rooted at a source site,
we suggest keeping one or more distance triples, for each triangle, that may help determine the Voronoi bisectors when one uses a mark-and-sweep geodesic algorithm to 
predict the multi-source distance field.
Second, rather than keep the distance itself at a mesh vertex, we use the squared distance to characterize the linear change of distance field restricted in a triangle, which is proved to induce an exact VD when the base surface reduces to a planar triangle mesh.
Specially, our algorithm also supports the Euclidean distance, which can 
handle thin-sheet models (e.g. leaf) and runs faster
than the traditional restricted Voronoi diagram~(RVD) algorithm. 
It is very extensible to deal with various variants of surface-based Voronoi diagrams including (1)~surface-based power diagram,
(2)~constrained Voronoi diagram with curve-type breaklines,
and (3)~curve-type generators. 
We conduct extensive experimental results to validate 
the ability to approximate the exact VD in different distance-driven scenarios.
\end{abstract}

%CCS
% <concept_id>10010147.10010371.10010372</concept_id>
% <concept_desc>Computing methodologies~Rendering</concept_desc>
% <concept_id>10010147.10010371.10010372.10010374</concept_id>
% <concept_desc>Computing methodologies~Ray tracing</concept_desc>

\ccsdesc[500]{Computing methodologies~Mesh geometry models}

%keywords
\keywords{digital geometry processing, geodesic distance,  geodesic Voronoi diagram, restricted Voronoi diagram}

\begin{teaserfigure}
  \centering
  \includegraphics[width=.99\textwidth]{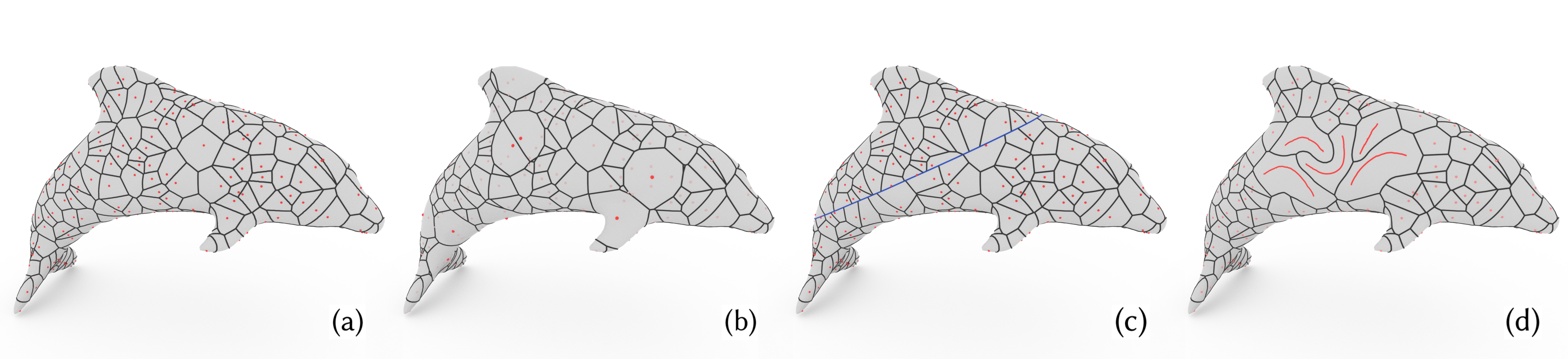}
  \vspace{-5mm}
  \caption{We propose a novel algorithm for computing surface-based Voronoi diagrams, without the need of an off-the-shelf Voronoi/Delaunay solver.
  Our algorithm supports an arbitrary geodesic algorithm to drive the partition; Specially, it also supports Euclidean distance~(a).
  Our algorithm is flexible enough to handle various versions of surface-based Voronoi diagrams. 
For example, it can be used to compute the surface-based power diagram~(b).
Furthermore, 
 it enables users to draw breaklines to prevent any Voronoi cell from crossing the breaklines~(c),
 and naturally supports curve segments as the source sites~(d).
  }
%\Description{Enjoying the baseball game from the third-base seats. Ichiro Suzuki preparing to bat.}
  \label{fig:teaser}
\end{teaserfigure}

\maketitle

\section{Introduction}
Partitioning the 2-manifold surface~$\mathcal{S}$
into regions based on the proximity to 
a given set of points~$\mathbf{P}=\{p_i\}_{i=1}^m$ 
is a fundamental operation in geometry processing.
Surface-based Voronoi diagram (VD) has many important applications ranging from
mesh quality improvement~\cite{levy2010p} to surface reconstruction~\cite{2016Fixed}, and can even mimic natural tessellation patterns~\cite{10.1145/3272127.3275072}.
It can be seen as the extension of Voronoi diagram (VD) in~$\mathbb{R}^2$ to the 2-manifold surface, while the difference is that geodesic distance serves as the metric to drive the computation of surface-based VDs. 
Although the study of VD in~$\mathbb{R}^2$ is relatively mature,
efficiently computing surface-based VDs is still a challenging task.

The challenges lie in several aspects.
First, in the past research,
the choice of geodesic algorithms~\cite{liu2017constructing,liu2010construction, wang2015intrinsic} is highly coupled with designing the scheme for computing surface-based VDs. 
Second, there are various versions of surface-based VDs but the known algorithms~\cite{yan2013efficient,yan2009isotropic,wang2020robustly} are not flexible  enough to provide an all-in-one solution. 
Finally, Euclidean distance is an approximate alternative to geodesic distance especially when
the sites are dense enough, which accounts for why the restricted Voronoi diagram~(RVD)~\cite{yan2009isotropic} is popular in the computer graphics community. However, it is not easy to adapt RVD to deal with a thin-sheet model because there are point pairs with very small spatial distance but long geodesic distance on these models.
% However, it is hard for the existing algorithms~\cite{yan2009isotropic} to own the two properties at the same time: (1)~the resulting surface partitioning is proved to be a natural extension of Euclidean VD, and (2)~the algorithm still works well on a thin-sheet model.

It is observed that the surface-based Voronoi diagram
can be extracted by finding the lower envelope of the $m$ distance fields,
each one rooted at a source site. 
By taking the change of a single-source distance field restricted in a triangle~$f$
as linear, the surface VD in~$f$, determined by multiple sites, can be computed by intersecting a collection of half-planes. Inspired by this observation, we propose a powerful algorithm, named {\em SurfaceVoronoi}, for computing surface-based Voronoi diagrams, without the need of an off-the-shelf Voronoi/Delaunay solver. 
The contributions are two-fold.
On one hand, we adapt the user-specified mark-and-sweep geodesic algorithm
such that each triangle keeps one or more distance triples that may help determine the Voronoi bisectors.
Note that even if a source cannot provide the closest distance to any vertex of the triangle, it may still provide the closest distance to an area within the triangle. 
Therefore, for the existing multi-source geodesic algorithms~\cite{moenning2004intrinsic} that only keep the closest source for each mesh vertex, it is impossible to infer the surface VD structure accurately. Additionally, rather than keep the distance itself at mesh vertices, we use the squared distance to define the linear change in a triangle, which facilitates
reporting the exact VD when the curved surface reduces to a planar triangle mesh. 

Our SurfaceVoronoi algorithm has many distinguished features. 
First, it can work with an arbitrary geodesic algorithm,
and even supports the Euclidean distance as the driving measure; See Fig.~\ref{fig:teaser}(a).
Depending on the geodesic-distance type, our algorithm can produce various kinds of diagrams including
exact geodesic Voronoi diagram (EGVD),
approximate geodesic Voronoi diagram (AGVD), and Euclidean distance based Voronoi diagram (EDBVD),
where
EDBVD can naturally 
handle thin-sheet models and runs faster
than the traditional restricted Voronoi diagram~(RVD).
Second, our algorithm is flexible enough to deal with various versions of surface-based Voronoi diagrams.
For example, it can be used to compute the surface-based power diagram (see Fig.~\ref{fig:teaser}(b)). 
Furthermore, it can handle the scenario where users draw some breaklines to prevent any Voronoi cell across the breaklines (see Fig.~\ref{fig:teaser}(c)). It even supports curve-type sites or face-interior sites (see Fig.~\ref{fig:teaser}(d)).
Finally, it could work on a base surface equipped with a non-uniform density field.

\section{Related Work}
\subsection{VD in Euclidean spaces}
Given a set of generators (also called sites) $\mathbf{P}=\{p_i\in \mathbb{R}^d\}_{i=1}^m $, the VD defines the partition of $\mathbb{R}^d$ into convex regions (may be unbounded) based on the
proximity to the generators,
where the site~$p_i$ dominates the subregion
\begin{equation}\label{eq:voronoi}
  \Omega_i = \{x \in \mathbb{R}^d  \ \big| \ \|x - p_i\|\leq \|x - p_j\|,\  \forall j\neq i\}.
\end{equation}
VD, as well as its dual (i.e., Delaunay triangulation), has been widely applied to many science and engineering fields~\cite{senechal1993spatial}, including computer graphics, image processing, robot navigation, computational chemistry, materials science, and climatology.
There is a large body of literature on computing VDs in computational geometry. 
Some typical computing paradigms include
the  divide-and-conquer scheme~\cite{shamos1975closest},
incremental construction~\cite{green1978computing},
and Fortune's sweep line~\cite{fortune1987sweepline}.
An interesting fact is that 
VD or Delaunay triangulation can be constructed by a lifting technique~\cite{fortune1995voronoi}.

\subsection{VD on 2-manifold surfaces}
The definition of VD can be extended to a general metric space different from Euclidean spaces~\cite{giblin2004formal}.
Construction of VDs on curved surfaces, particularly non-differentiable polyhedral surfaces, is of great significance in digital geometry processing. 
% However, many properties on the smooth manifold do not hold
% any more on a piecewise linear triangle mesh, the extension from Euclidean VD to surface-based VD
% still confronts many unresolved challenges.
% The technical challenges are mainly two-fold.
% First, the estimation of geodesic distances is not so simple as straight-line distances.
% Second, finding the geodesic bisectors is non-trivial since the sites cannot be unfolded onto a plane.
% Roughly speaking, existing research works on computing surface-based VDs can be divided into two kinds. 
% The first kind of works aims to 
% trace the bisectors by employing the window propagation scheme.
% For example, 
% Liu et al.~\cite{liu2010construction} studied the analytic structure of iso-contours, bisectors, and Voronoi diagrams on a triangular mesh and further proposed 
% a geodesic VD algorithm by 
% considering the running scheme of MMP's algorithm~\cite{mitchell1987discrete}. 
% The intrinsic Delaunay triangulation 
% can thus be induced from geodesic VD~\cite{liu2017constructing}.
% Xu et al.~\cite{xu2014polyline} proposed an algorithm for constructing the GVD with polyline generators by utilizing 
% the window propagation scheme.
% Later, Qin et al.~\cite{qin2017fast} also presented
% an algorithm for constructing Voronoi diagrams based on the 
% window propagation mechanism of VTP.
% Although exact geodesic algorithms, e.g., MMP, VTP, can compute Voronoi diagrams accurately, they are time-consuming to obtain exact geodesic distances which
% become stumbling block of constructing GVDs and farther practical applications.
Roughly speaking, the methods used for computing surface-based Voronoi diagrams 
can be categorized into three groups, including (1)~operating on the mesh surface  directly based on geodesic distance~\cite{liu2010construction, wang2015intrinsic}, (2)~operating on the parametrization plane~\cite{Rong2010HyperbolicCV} and (3)~operating in the 3D Euclidean space~\cite{yan2009isotropic,yan2014low,wang2020robustly} with the help of a 3D Voronoi/Delaunay solver. 
%However, neither of them seems perfect.
Generally speaking, geodesic distance based approaches are time-consuming and coupled with the scheme for computing surface-based VDs. %and have to employ the window propagation scheme.coupled with designing the scheme for computing surface-based VDs.
For example, 
Liu et al.~\shortcite{liu2010construction} studied the analytic structure of iso-contours, bisectors, and Voronoi diagrams on a triangular mesh and further proposed 
a geodesic Voronoi diagram (GVD) algorithm based on 
the window propagation scheme of MMP's algorithm~\cite{mitchell1987discrete}. 
% The intrinsic Delaunay triangulation 
% can thus be induced from geodesic VD~\cite{liu2017constructing}.
There are some other GVD approaches that are highly coupled with the 
window propagation mechanism~\cite{xu2014polyline,qin2017fast}.
Parametrization-based approaches~\cite{Rong2010HyperbolicCV,Alliez2005CentroidalVD} need to
precompute a global parametrization and thus 
suffer from numerical issues and topological difficulties.
By substituting Euclidean distances for geodesic distances,
the restricted VD (RVD)~\cite{yan2009isotropic,yan2014low,wang2020robustly}
runs many times faster, but for models with thin parts, it may violate the property of “one site, one region”. 
Only when the sites are enough and dense,
can RVD be taken as an approximate alternative to GVD.
Later, Yan et al.~\shortcite{yan2014low} proposed localized RVD (LRVD), first inferring neighboring relationship
and then partitioning the surface, to 
handle thin-plate models,
but LRVD does not explicitly address poor triangulation quality.
Wang et al.~\shortcite{wang2020robustly} 
gave a fast post-processing technique for fixing the problematic 
RVD cells by enforcing the property of ``one site, one region''. 
Herholz et al.~\shortcite{https://doi.org/10.1111/cgf.13116}
proposed diffusion diagrams based on heat diffusion.
Their algorithm is only equivalent to back-substitution if factorization of the system matrix is done in a preprocess.

In this paper, we propose an all-in-one solution
to decouple the geodesic solver from the computation of surface VDs. 
Furthermore, we bridge the gap between geodesic VDs and Euclidean VDs. 
% Specially, our algorithm supports straight-line distance as well; In this situation, it runs as fast as the RVD algorithm but our algorithm can naturally 
% handle thin-sheet models.

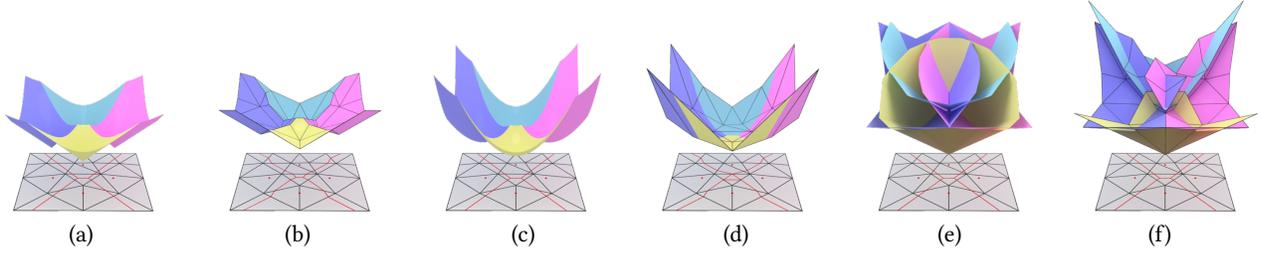
\begin{figure*}[ht]
%\vspace{-3.0mm}
	\centering
\begin{overpic}
[width=\linewidth]{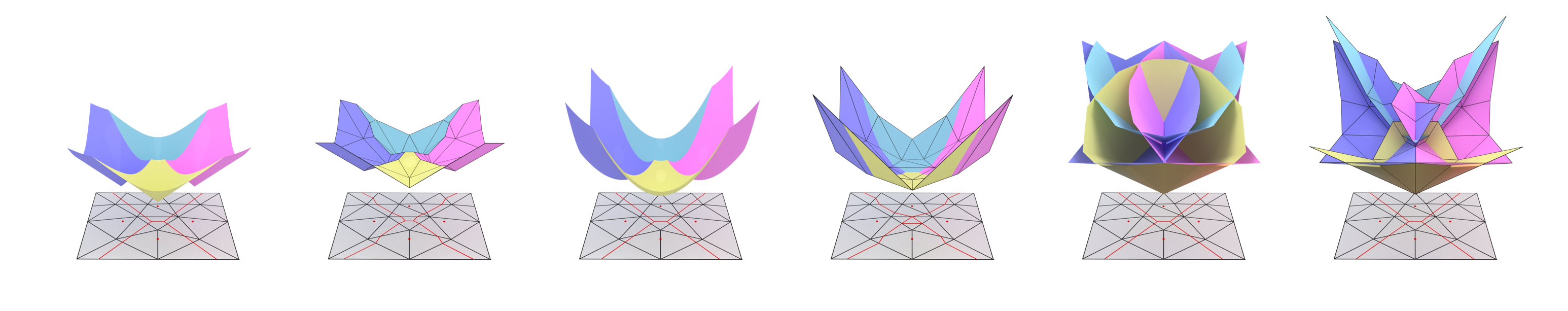}
\put(9,1){(a)}
\put(25,1){(b)}
\put(41.8,1){(c)}
\put(57.5,1){(d)}
\put(73.4,1){(e)}
\put(89,1){(f)}
\end{overpic}
\vspace{-5mm}
\caption{The VD based on lifting distance fields.
(a)~In 2D, the VD can be obtained by finding the lower envelope of a set of distance fields $\{\mathbf{D}_i\}_{i=1}^m$, each of which is a cone rooted at a site.
(b)~It can be easily extended to polygonal meshes. 
By finding the lower envelope of $m$ discrete distance fields, 
one can get an approximate VD. 
%followed by and tracing the GVD triangle by triangle. 
(c)~If we use the squared distance to define a continuous parabolic surface rooted at each site, the lower envelope can also report the VD.
(d)~It is hard to get an accurate VD from the multi-source distance field.
(e)~Interestingly, the lower envelope of the $m$ discrete squared-distance fields is able to report the exact VD when the base surface is reduced to a planar surface (we have proved the observation in this paper).
(f)~We further propose an {over-propagation} technique 
that can be used with any mark-and-sweep algorithm.
The fact is that each triangle keeps one or more necessary linear distance fields, which may help determine the Voronoi bisectors. 
}
\label{fig:insight}
	\vspace{-3mm}
\end{figure*}

\section{Overview}
%\subsection{Problem statement}
% Like the Voronoi diagram~(VD) in the 2D Euclidean space~$\mathbb{R}^2$,
% the VD on a 2-manifold surface $\mathcal{S}$ can be defined likewise.
% The main difference lies in the way that the distance between two points is measured. Given a set of sites $\mathbf{P}=\{p_i\}_{i=1}^m$ belonging to~$\mathcal{S}$, the region dominated by $p_i$ is defined as follows:
% \begin{displaymath}
% \Omega_i = \left\{\mathbf{x}\in \mathcal{S}\,\big |\, \mathbf{d}_g(\mathbf{x},p_i)\leq\mathbf{d}_g(\mathbf{x},p_j),\forall i\neq j\right\},
% \end{displaymath}
% where $\mathbf{d}_g(\cdot,\cdot)$ is the operator of taking geodesic distance.

% VDs in Euclidean spaces have been widely studied
% and many handy solvers are available,
% but computing geodesic Voronoi diagrams~(GVD) on a curved surface~$\mathcal{S}$ in 3D is not easy. 
% The technical challenges are mainly two-fold.
% First, the estimation of geodesic distances is not so simple as straight-line distances.
% Second, finding the geodesic bisectors is non-trivial since the sites cannot be unfolded onto a plane.
We first consider a toy example with four sites specified on a 2D plane; See Fig.~\ref{fig:insight}.
For each site $p_i$, the distance field rooted at~$p_i$ can be visualized as an inverse cone, and the lower envelope of the four cones defines the multiple-source distance field as well as the two-dimensional VD; See Fig.~\ref{fig:insight}(a).
It can be seen that the multiple-source distance field has sharp ridges (whose projection induces the VD structure) where the distance change is far from being linear.

Now suppose that the input surface~$\mathcal{S}$ is a planar or curved triangle surface. 
Let 
\begin{equation}
    \mathbf{D}_i=\left\{d_i^{(1)},d_i^{(2)},\cdots,d_i^{(j)},\cdots,d_i^{(n)}\right\}
\end{equation} 
be the $i$-th discrete distance field that keeps the distances at vertices,
and $n$ is the total number of vertices. 
By taking the change of a single-source distance field to be linear inside a triangle, 
the vector of $\mathbf{D}_i$ defines a piecewisely-linear approximation of the real distance field using~$n$ scalar values.
The lower-envelope technique can be thus extended to polygonal surfaces
by minimizing the group of piecewisely-linear geodesic distance fields,
triangle by triangle; See Fig.~\ref{fig:insight}(b). 
In fact, a linear distance field defined in a triangle
can be understood as a half-plane over the base triangle.
For each triangle, the lower envelope,
determined by $m$ sites, can be obtained by intersecting $m$ half-planes.

Lifting~\cite{fortune1995voronoi} is a well-known technique that 
maps $(x,y)$ to $(x,y,x^2+y^2)$. Obviously,
if we define a parabolic surface rooted at each site with squared distance, the lower envelope can also report the VD; See Fig.~\ref{fig:insight}(c). %In fact, the lifting mapping has many choices.

There are many research approaches~\cite{moenning2004intrinsic} 
that suggest computing the surface-based VDs by running the multi-source
geodesic distance algorithm that only keeps a minimum distance for each mesh vertex: 
\begin{displaymath}
\mathbf{D}_\text{min} =\left\{d_j^\text{min}:\,\text{the~distance~from~$v_j$~to~the~nearest~site} \right\}_{j=1}^n.
\end{displaymath}
But it is hard for them to extract an accurate VD from $\mathbf{D}_\text{min}.$
% The final approximate partitioning structure can be extracted by connecting {\em edge pivots} and {\em triangle pivots}.
% An edge pivot $q_{v_iv_j}$ on the edge $v_iv_j$ is found by solving the equation
% \begin{displaymath}
% \mathbf{D}_\text{min}(v_i)+\|q_{v_iv_j}v_i\|=\mathbf{D}_\text{min}(v_j)+\|q_{v_iv_j}v_j\|,
% \end{displaymath}
% where $\mathbf{D}_\text{min}(v_i)$ and $\mathbf{D}_\text{min}(v_j)$ are given by different sources.
% Similarly, a triangle pivot $q_f$ in the triangle $v_iv_jv_k$ is found by solving the min-max problem:
% \begin{displaymath}
% q_f=\min_{q\in f}\max\{\mathbf{D}_\text{min}(v_i)+\|qv_i\|,
% \mathbf{D}_\text{min}(v_j)+\|qv_j\|,
% \mathbf{D}_\text{min}(v_k)+\|qv_k\|\},
% \end{displaymath}
% where $\mathbf{D}_\text{min}(v_i),\mathbf{D}_\text{min}(v_j),\mathbf{D}_\text{min}(v_k)$ are given by three different sources.
As Fig.~\ref{fig:insight}(d) shows, the traced bisectors are visually zigzag, due to 
the fact that $\mathbf{D}_\text{min}$, 
keeping only a minimum distance for each mesh vertex,
is not informative enough
to trace the VD structure. 
% , unlike the multiple-source distance field $\mathbf{D}_\text{min}$,
% based on whether geodesic distance or squared distance, lacks the complete knowledge of the non-smooth distance change in the zone along bisectors.

% In fact, the resulting multi-source distance field 
% is not informative enough to manifest the non-smooth change of the lower envelop of the $m$ distance field, making the extracted VD zigzag; See Fig.~\ref{fig:insight}(d).
% The most popular way is to extract the GVD from the multiple-source distance field:

In this paper, we propose to compute surface-based VDs
based on two novel techniques. 
First, when inferring the lower envelope for a triangle,
we simply replace the distance value at a vertex 
by the squared distance. 
The resulting VD is exact when $\mathcal{S}$ reduces to a planar surface; 
See Fig.~\ref{fig:insight}(e),
which is much different from the situation of Fig.~\ref{fig:insight}(b).
In Section~\ref{subsec:squared},
we prove the observation
that squared-distance induces a natural extension of the VD 
from 2D to a curved surface.
% have an interesting observation that
% the lower envelope of the $m$ discrete squared-distance fields is able to report the exact VD when the base surface reduces to a planar surface (see Fig.~\ref{fig:insight}(e)), which shall be proved in Section~\ref{subsec:squared}. Based on this observation,
% squared-distance induces a natural extension of the VD 
% from 2D to a curved surface, without any additional computational amount. 
Second, it is obvious that computing a full distance field for every source site is time-consuming and unnecessary,
so we propose an {\em over-propagation} distance field, where the distance field rooted at~$p_i$ is swept outward until it does not help determine the surface-based VD at all.
The key strategy is to introduce a comparison rule between two triples
(each triple defines a linear distance field in a triangle). 
It can be seen from Fig.~\ref{fig:insight}(f) that the resulting diagram  has no difference from Fig.~\ref{fig:insight}(e), but 
the sweeping area is shrunk, greatly diminishing
the computational cost.
We shall elaborate on the two techniques in the next section. 
% The theme of this paper is to study how to compute GVDs based on over-propagation
% distance fields.

\paragraph{Remark}

As Fig.~\ref{fig:insight}(e) shows,
the assumption of squared-distance linearity 
coincidentally produces an exact VD when $\mathcal{S}$ reduces to a planar surface;
See Section~\ref{subsec:squared}.
But we must point out that the assumption does not apply to a curved surface. 
Despite this, the linearity assumption in terms of squared distance
is still better than Euclidean distance; See Section~\ref{sec:Evaluation}.
Besides, the linearity assumption holds only if 
the triangles are small enough because the geodesic distance field satisfies the Lipschitz continuity (note that the gradient norm cannot exceed 1).

\section{Algorithm}
To facilitate the explanation,
we first introduce the over-propagation geodesic field
in Section~\ref{subsec:over-propagation}, and then justify how to lift the geodesic distance to the squared distance
in Section~\ref{subsec:squared}. 
%the linear approximation of a distance field using the squared-distance.
\subsection{Over-propagation}
\label{subsec:over-propagation}
Essentially, the surface-based VD
is formed due to the competition between different sites,
i.e.,
a point $x\in\mathcal{S}$ belongs to $p_i$ 
if and only if $p_i$ can provide a shorter distance to $x$ than the other sites. 
Suppose that we have an off-the-shelf mark-and-sweep distance
field algorithm (e.g., the fast marching method or the window based exact geodesic algorithms).
There are two important aspects when using the distance solver to compute surface-based VDs:
(1)~how to maintain the sweeping of $\{\mathbf{D}_i\}_{i=1}^m$ collaboratively and constrain each of them to sweep across an as-small-as-possible area,
and (2)~how to quickly fuse the distance fields to get the lower envelope that defines the surface-based VD.

 \begin{algorithm}[t]
 \SetAlgoNoLine
 \KwIn{A curved polygonal surface~$\mathcal{S}$ and a set of source points $\mathbf{P}=\{p_i\}_{i=1}^m$ on~$\mathcal{S}$.}
 \KwOut{Over-propagation distance field $\mathbf{D}_\text{min}$ rooted $\mathbf{P}$.}
 \ForEach{triangle $f$}
 {
 Associate $f$ with an empty half-plane list ($\bigstar$)\;
 }
 Initialize an empty priority queue $\mathbf{Q}$ for distance propagation\;
  \ForEach{source point $p\in\{p_i\}_{i=1}^m$}
 {
 Push a distance propagation event rooted at $p$ to $\mathbf{Q}$\;
 }
 \While{$\mathbf{Q}$ is not empty}
 {
Take out the top-priority event $evt$\;
Extract the source point $p$ encoded in $evt$\;
\quad//When $evt$ arrives at a mesh vertex $v$, we need to check each of $v$'s incident faces\;
Extract the triangle $f$ that $evt$ arrives at\;
Check if $evt$ can update the distance at one of the three vertices of $f$ ($\bigstar$)\;
\If {$p$ is a {contributing} source point for $f$}
{
Append $p$ to the source-point list of $f$ ($\bigstar$)\;
Allow $evt$ to generate new events of next generation in $f$ and push them into $\mathbf{Q}$\; 
}
}
 \caption{Using a typical mark-and-sweep algorithm to perform over-propagation, where the differences from the traditional multi-source sweep algorithm are highlighted with  ``$\bigstar$''.}
 \label{alg:mark-and-sweep}
  \end{algorithm}

 \iffalse
 \begin{figure}[t]
%\vspace{-3.0mm}
	\centering
\begin{overpic}
[width=.7\linewidth]{figs/fastmarching.png}
\end{overpic}
\caption{The fast marching method is a typical mark-and-sweep algorithm. \PF{not mentioned in paper}}
\label{FIG:FMM}
	\vspace{-3mm}
\end{figure}
\fi
%\subsection{Mark-and-sweep distance propagation algorithms} %\label{subsec:Mark-and-sweep}
\paragraph{Mark-and-sweep}
There are many mark-and-sweep algorithms for computing a distance field or finding the shortest path on a polygonal surface~$\mathcal{S}$, such as~\cite{kimmel1998computing,du2021vertex,qin2017fast,xin2009improving}. 
Although being different in algorithm details, they share a similar algorithmic paradigm - propagating from near to far and prioritizing near-source distance propagation events,
where each event contains a pair, i.e., the source point and the triangle,
and also a value about the closest distance between them.
The priority of events as well as the morphology of wavefronts is maintained by a priority queue, such that the events in the priority queue
are processed according to the closest distances encoded by the events.
When the source point of the current event cannot provide the closest distance to any vertex of the current triangle, the propagation process of the current source point on this triangle stops.
The whole mark-and-sweep algorithm terminates when 
% At the end of a mark-and-sweep algorithm, each point on~$\mathcal{S}$ (typically every vertex) 
each vertex finds its nearest source point, resulting in a multi-source distance field.
However, such a distance field 
is not informative enough to accurately trace the bisectors. 
Therefore, we keep all source points that contribute to each complete triangle (including its interior points), not just the vertices of the triangle. Details will be discussed below.

% Such a distance field is not informative since for a point $q$ inside a triangle across the GVD, how far it is from the sources and which source gives the distance to $q$ are unknown yet.
% Based on such a distance field, the triangle-by-triangle extracted GVD is very inaccurate. 

\begin{figure}[t]
%\vspace{-3.0mm}
	\centering
\begin{overpic}
[width=.99\linewidth]{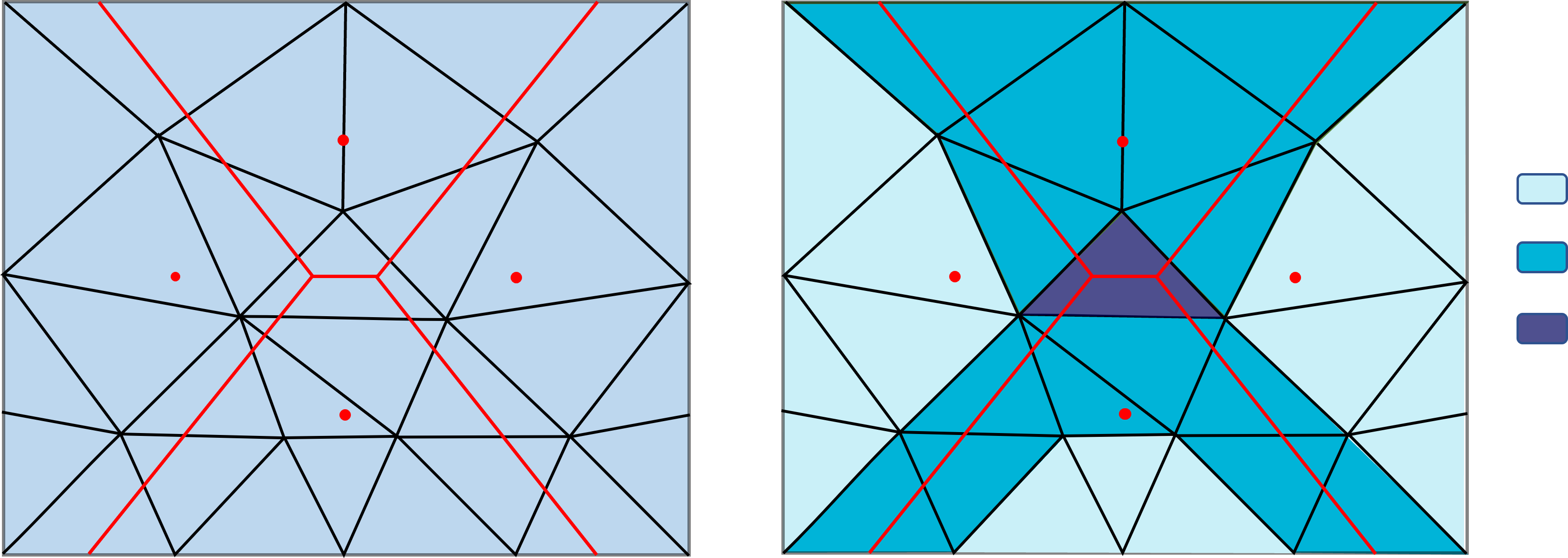}
\put(100,22.5){~(1)}
\put(100,18){~(2)}
\put(100,13.6){~(4)}
\end{overpic}
\caption{Left: a planar mesh and the VD w.r.t. four sites.
Right: for each triangle, we visualize how many distance fields (sources) really contribute to the triangle in a color-coding style.
}
\label{FIG:SourceNum}
	\vspace{-3mm}
\end{figure}

\paragraph{Filtering rule in over-propagation}
A source point is not allowed to 
continue propagating 
across a triangle $f\triangleq v_1v_2v_3$, only if it is defeated
by other source points in $f$. %in every wavefront triangle.
%Theoretically speaking, 
In other words, a source point $p$ is said to be a {\em contributing} site to the triangle $f$, if $p$ can provide the shortest distance for at least one point in $f$. 
Under the assumption that 
the distance field given by a single source is
approximately linear in every triangle,
$p_i$ is defeated by $p_j$ in~$f$, only if
\begin{displaymath}
d_j^{(1)}<d_i^{(1)},
d_j^{(2)}<d_i^{(2)},
d_j^{(3)}<d_i^{(3)},
\end{displaymath}
where $d_i^{(1)},d_i^{(2)},d_i^{(3)}$ are the three distances at $v_1,v_2,v_3$ given by $p_i$
while $d_j^{(1)},d_j^{(2)},d_j^{(3)}$ are given by $p_j$.
%Under the assumption of linear change, 
$p$ is a contributing site to~$f$ if 
% $p$ is the earliest one that arrives at $f$ among all the source points or
it cannot be defeated by any other source point in $f$'s contributing-source list.
As Algorithm~\ref{alg:mark-and-sweep} indicates, existing mark-and-sweep geodesic algorithm 
can be easily used for over-propagation
by organizing all the propagation events into a prioritized queue.
In Algorithm~\ref{alg:mark-and-sweep}, the differences from the traditional multi-source sweep algorithm are highlighted with  ``$\bigstar$''.
In Fig.~\ref{FIG:SourceNum},
we visualize distance triples (or equivalently source sites) that really contribute to the triangle at the end of the Algorithm~\ref{alg:mark-and-sweep}.
% Fig.~\ref{??}(a) shows an example of over-propagation distance field, where
% we paint the triangles in different colors depending on the number of viable source points. Fig.~\ref{??}(b) shows a traditional multi-source distance field, where each vertex is marked with the source point that defines the shortest distance. 

%\SQ{a figure please}\XR{Doen.}

% \begin{figure}[h]
% %\vspace{-3.0mm}
% 	\centering
% \begin{overpic}
% [width=.99\linewidth]{figs/SingleFace2.png}
% \put(15,-5.0){~(a)Ours}
% \put(60,-5.0){~(b)Diffusion Diagrams}
% \end{overpic}
% \vspace{2mm}
% \caption{single face}
% \label{FIG:SourceNum}
	
% \end{figure}
%\vspace{-4mm}
%\subsection{Finding the lower envelope of half-planes}
%\label{subsec:half-plane}
\paragraph{Finding the lower envelope of half-planes}
Suppose that $\triangle v_1v_2v_3$ is one of the triangles in the mesh.
At the end of over-propagation, we assume that $p_1,p_2,\cdots,p_K\in \mathbf{P}$ really contribute to $\triangle v_1v_2v_3$.
Each site in~$\mathbf{P}$ gives a distance triple that encodes a linear field in terms of geodesic distance. (Note that we will discuss the possibility of representing the linearity w.r.t. the squared distance.)
Without loss of generality, we place $\triangle v_1v_2v_3$ on a 2D plane, by a rigid transform, such that the vertex $v_j$ is located at $(x_j,y_j),j=1,2,3.$
$p_i$'s distance field constrained in $\triangle v_1v_2v_3$
can be deemed to change linearly. We can use a half-plane $\pi_i$ to define the linear approximation, where $\pi_i$ passes through the following three points:
\begin{displaymath}
\left(x_1,y_1,\mathbf{d}_g(p_i,v_1)\right),
\left(x_2,y_2,\mathbf{d}_g(p_i,v_2)\right),
\left(x_3,y_3,\mathbf{d}_g(p_i,v_3)\right),
\end{displaymath}
where $d_g(p_i,v_j)$ means the distance between the source point $p_i$ and the triangle vertex $v_j$.

\begin{figure}[t]
%\vspace{-3.0mm}
	\centering
\begin{overpic}
[width=.99\linewidth]{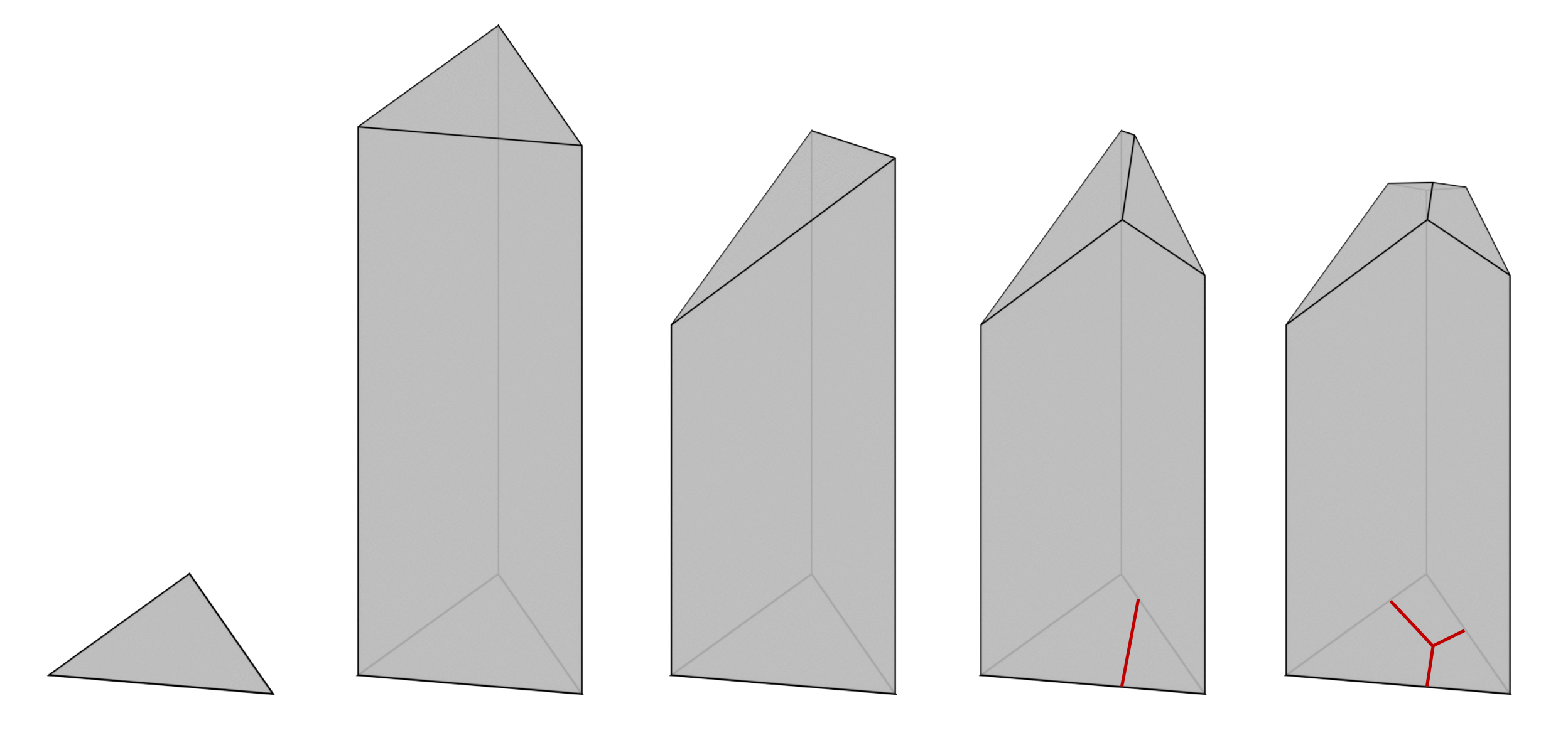}
\put(8,-1){(a)}
\put(28,-1){(b)}
\put(48.5,-1){(c)}
\put(67.5,-1){(d)}
\put(87,-1){(e)}
\end{overpic}
\caption{
We use a half-plane to encode the linear change
of a distance field inside a triangle.
%When there are many distance fields arriving the triangle, 
During the mark-and-sweep process,
the lower envelope over the triangle base can be obtained by incremental half-plane cutting.
The VD restricted in the triangle can be traced accordingly. 
}
\label{FIG:partition}
	\vspace{-4mm}
\end{figure}
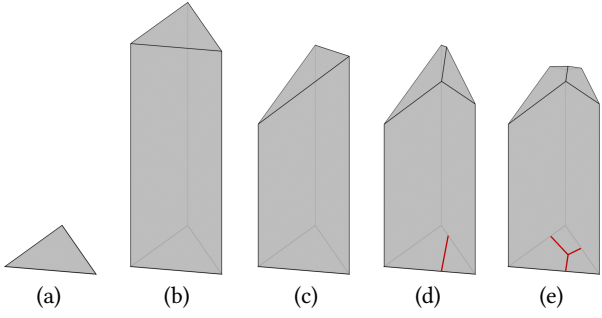

The next task is to find the lower envelope of $\{\pi_i\}_{i=1}^K$.
Our basic idea is to construct an infinite triangular prism with $\triangle v_1v_2v_3$ being the base and then incrementally cut 
the volume by $\{\pi_i\}_{i=1}^K$, like that achieved in~\cite{10.1145/3450626.3459870}. 
% For consideration of efficacy, we implemented the operation of incremental half-plane cutting without using existing numerical packages. 
We denote $d_i^{(1)}\triangleq \mathbf{d}_g(p_i,v_1),d_i^{(2)}\triangleq \mathbf{d}_g(p_i,v_2),d_i^{(3)}\triangleq \mathbf{d}_g(p_i,v_3)$.
% \paragraph{Half-plane.}
% Let $d_i^{(1)},d_i^{(2)},d_i^{(3)}$ be respectively the three distances at $v_1,v_2,v_3$, given by the source $p_i$.
Since each half-plane $\pi_i$ cannot be vertical, we adopt the following representation form:
\begin{displaymath}
d=a_ix+b_iy+c_i,
\end{displaymath}
where 
\begin{displaymath}
\begin{pmatrix}
a_i\\b_i\\c_i
\end{pmatrix}=\begin{pmatrix}
x_1&y_1&1\\
x_2&y_2&1\\
x_3&y_3&1\\
\end{pmatrix}^{-1}
\begin{pmatrix}
d_i^{(1)}\\
d_i^{(2)}\\
d_i^{(3)}
\end{pmatrix}.
\end{displaymath}
Given a triple $(x,y,d)$, the triple is on the upper side of $\pi_i$
if 
\begin{displaymath}
d\geq a_ix+b_iy+c_i,
\end{displaymath}
and on the lower side if 
\begin{displaymath}
d< a_ix+b_iy+c_i.
\end{displaymath}
Fig.~\ref{FIG:partition} shows the incremental plane cutting process. The VD restricted in the triangle can be reported at the same time. 
\paragraph{Implementation}
We give the pseudo-code of incremental half-plane cutting procedure in Algorithm~\ref{alg:cutting}.
Each time
when a new half-plane $\pi$ comes, we will kill the existing vertices (of the up-to-date lower envelope) that are above $\pi$.
Next, we identify those edges that pass through $\pi$.
% the check the vertices to see if they are still alive.
% We erase the vertex $v_i$ if $v_i$ is above $\pi$. 
% After that, we check the positional relationship between the alive edges and $\pi$.
Suppose we are considering the edge $e=v_1v_2$.
%(3) One is above $\pi$ and the other is under $\pi$. 
If $\pi(v_1)=d_{v_1}-(ax_{v_1}+by_{v_1}+c)>0$ 
while $\pi(v_2)=d_{v_2}-(ax_{v_2}+by_{v_2}+c)<0,$ 
we need to add a new vertex onto the up-to-date lower envelope:
%We append the following new vertex to the back of the vertex array: 
\begin{displaymath}
v=\frac{\pi(v_1)}{\pi(v_1)-\pi(v_2)}v_2-\frac{\pi(v_2)}{\pi(v_1)-\pi(v_2)}v_1.
\end{displaymath}
The bounding edges of the envelope 
also need to be updated at the same time but the bounding facets do not. In addition, for consideration of run-time performance, we maintain the lower envelope for the triangle $f$ only when the second triple (linear distance field) comes to $f$
and the new triple has a chance to survive. 
Let $K$ be the number of half-planes kept in the current triangle~$f$. 
The number of vertices/edges/faces of the lower envelope is $O(K)$.
The total time cost is bounded by $O(K^2)$.

% Once the new vertex is generated, we update the lower envelope based on the following three steps. 
% First, we remove $v_1$ from the alive vertex set.
% Second, we relate $e$'s two incident half-planes as well as the new half-plane $\pi$ to $v$.
% Finally, we keep $e$ still alive while updating~$e$'s endpoint $v_1$ to $v$.
% At the end of half-plane cutting, the edge set exactly reports the VD constrained in~$\triangle v_1v_2v_3$.
% We summarize the incremental half-plane cutting procedure in Algorithm~\ref{alg:cutting}.

 \begin{algorithm}[t]
 \SetAlgoNoLine
 \KwIn{$\triangle v_1v_2v_3$ and a set of half-planes~$\{\pi_i\}_{i=1}^K$.}
 \KwOut{Partition of $\triangle v_1v_2v_3$ based on the lower envelope of $\{\pi_i\}_{i=1}^K$.}
 Initialize a vertical prism truncated by $h=d_\text{max}$ and $h=-d_\text{max}$\;
 \ForEach {$\pi\in \{\pi_i\}_{i=1}^K$}
 {
Screen out vertices above $\pi$\;
Screen out useless edges\;
\ForEach {edge $e$ crossing $\pi$}
 {
 Generate a new vertex\;
 Update $e$'s endpoint\;
 }
 }
 Output the alive edges as the VD in $\triangle v_1v_2v_3$\;
 \caption{Incremental half-plane cutting}
 \label{alg:cutting}
 \end{algorithm}

\subsection{Compatibility, consistence and connectedness}
\begin{figure}[t]
%\vspace{-3.0mm}
	\centering
\begin{overpic}
[width=.95\linewidth]{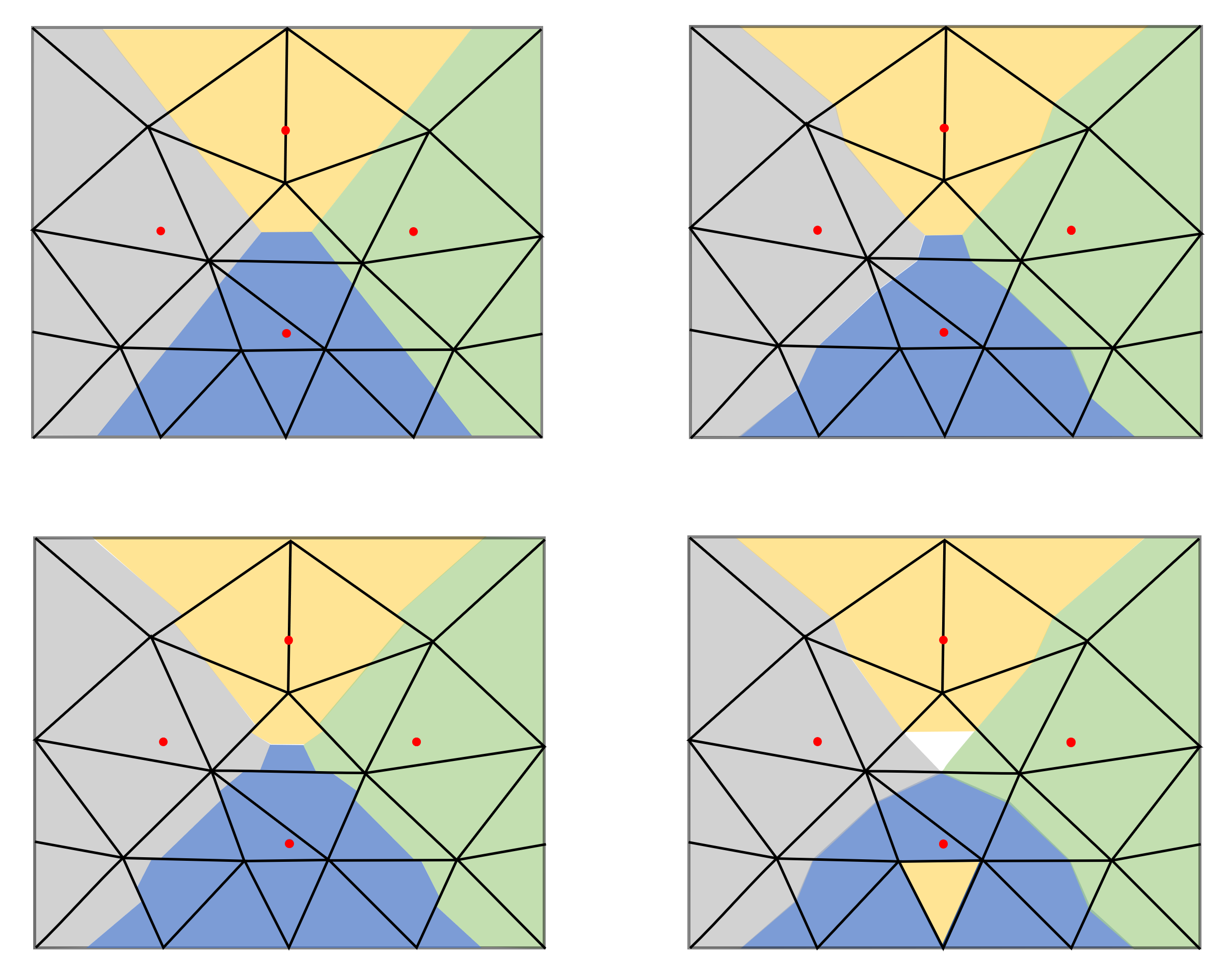}
\put(6,40){(a)~An acceptable VD}
\put(55,40){(b)~Not compatible to 2D VD}
\put(3,-2){(c)~Inconsistent across edges}
\put(57,-2){(d)~Disconnected regions}
\end{overpic}
\caption{For the applications, 
one hopes to obtain an acceptable VD 
that owns the following features:
(1)~it naturally reports the exact VD when the base surface reduces to a
convex 2D domain,
(2)~the VD must be consistent across edges, i.e., passes through the same point of an edge for the two faces sharing the edge,
and (3)~no site is allowed to dominate multiple disconnected regions.
In this figure, (a) shows an acceptable VD while (b,c,d) respectively show three situations that violate the three requirements. 
}
\label{FIG:Compatibility}
	\vspace{-3mm}
\end{figure}
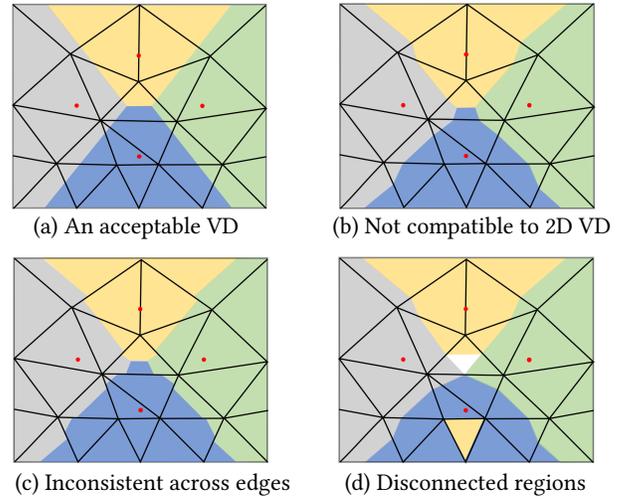
It seems that the past research
lacks indicators to evaluate the quality of a surface-based VD. 
We propose three indicators to characterize if a surface-based VD is valid (see Fig.~\ref{FIG:Compatibility}(a)). 
First, we hope that the resulting VD partition
has no difference from the exact VD when the base surface becomes a
convex 2D domain.
If some surface VD algorithms report a partition in Fig.~\ref{FIG:Compatibility}(b),
it is not {\em compatible} with the 2D Voronoi diagram.
Second, we hope that the extracted VD does not include breakpoints (see Fig.~\ref{FIG:Compatibility}(c)) at mesh edges,
i.e., for each mesh edge, the entry point of the VD
must be identical to the exit point.
We name this requirement {\em consistence}.
Finally, we hope that each site dominates a connected region (may not be topologically equivalent to a disk),
and all the regions form a covering of the whole surface.
In this case, such surface VDs own {\em connectedness}.
Fig.~\ref{FIG:Compatibility}(d) shows an example 
where a site may dominate two or more regions (see the regions colored in yellow). 
Furthermore, the occurrence of an ownerless region
also violates the requirement of connectedness; See the white region in Fig.~\ref{FIG:Compatibility}(d).

Obviously, our over-propagation algorithm inherits the spirit of minimizing $m$ continuous and piecewisely-linear distance fields.
The resulting distance field, i.e., the lower envelope of the $m$ distance fields,
is also continuous, which guarantees the consistence.
Besides, our algorithm is equipped with a mark-and-sweep distance solver, 
and thus each site detects its dominating area without destroying its connectivity,
which guarantees the connectedness. In the following, we propose to use squared distance to define half-planes, which enables to further own the feature of compatibility. 
\iffalse
\paragraph{Consistence and connectedness naturally guaranteed} 
Recall that our over-propagation distance field keeps the distances contributed by each separate site~$p$ in its nature - the only difference lies in that we stop the propagation for the triangles whose distances cannot be provided by~$p$; See Fig.~\ref{??}.
Furthermore, we take the change in each triangle to be linear. 
Obviously, the piecewise linear distance field given by $p$ is continuous. 
The consistence is immediately verified based on the continuity. 
In addition, we assume that the given geodesic algorithm has a mark-and-sweep style, sweeping from near to far, which naturally guarantees the connectedness.

\paragraph{Compatibility not guaranteed} 
The inset figure shows a single triangle, the coordinates of whose three vertices are \SQ{???}.
The triangle contains two sites \SQ{?} and \SQ{?}. 
The two half-planes are \SQ{?} and \SQ{?}.
Their intersection line, after being projected, becomes ??,
which is obviously different from the real bisector \SQ{?}.
Therefore, the surface-based VD reported by our algorithm is different from the traditional 2D Voronoi diagram.
We shall tackle this issue in the next section. 
\fi
\subsection{Using squared distance to define half-planes}
\label{subsec:squared}
\begin{definition}
We assume that $\triangle v_1v_2v_3$ and a point $p(x_p,y_p)$ are in the xy-plane ($p$ may be lying outside $\triangle v_1v_2v_3$), where the coordinates of the three vertices are $v_1(x_1,y_1),v_2(x_2,y_2),v_3(x_3,y_3)$. We define $\Pi_{v_1v_2v_3}^p\in\mathbb{R}^3$ to be the plane through the following three points:
\begin{displaymath}
(x_1,y_1,\|p-v_1\|^2),(x_2,y_2,\|p-v_2\|^2),(x_3,y_3,\|p-v_3\|^2),
\end{displaymath}
and $z(x,y;\Pi_{v_1v_2v_3}^p)$ to be the height at $(x,y)$.
\end{definition}

\begin{definition}
We assume that $\mathcal{P}:z=x^2+y^2$ be the parabolic surface.
Denote $\Pi_{\mathcal{P}}^p$ as the tangent plane at $p$'s lifted point~$(x_p,y_p,x_p^2+y_p^2)$,
i.e., $z=2x_p x + 2y_p y - x_p^2-y_p^2.$
Similarly, we use $z(x,y;\Pi_{\mathcal{P}}^p)$ to denote the height at $(x,y)$.
\end{definition}

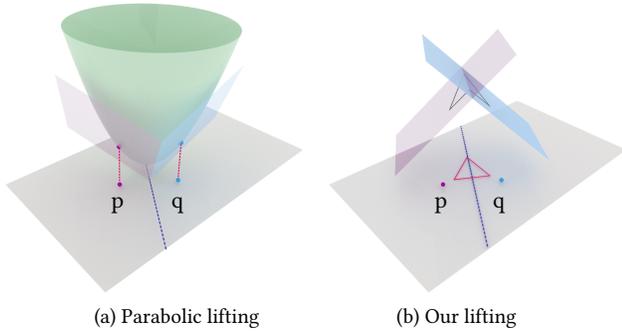
\begin{figure}[t]
  \begin{center}
  \begin{overpic}
[width=\linewidth]{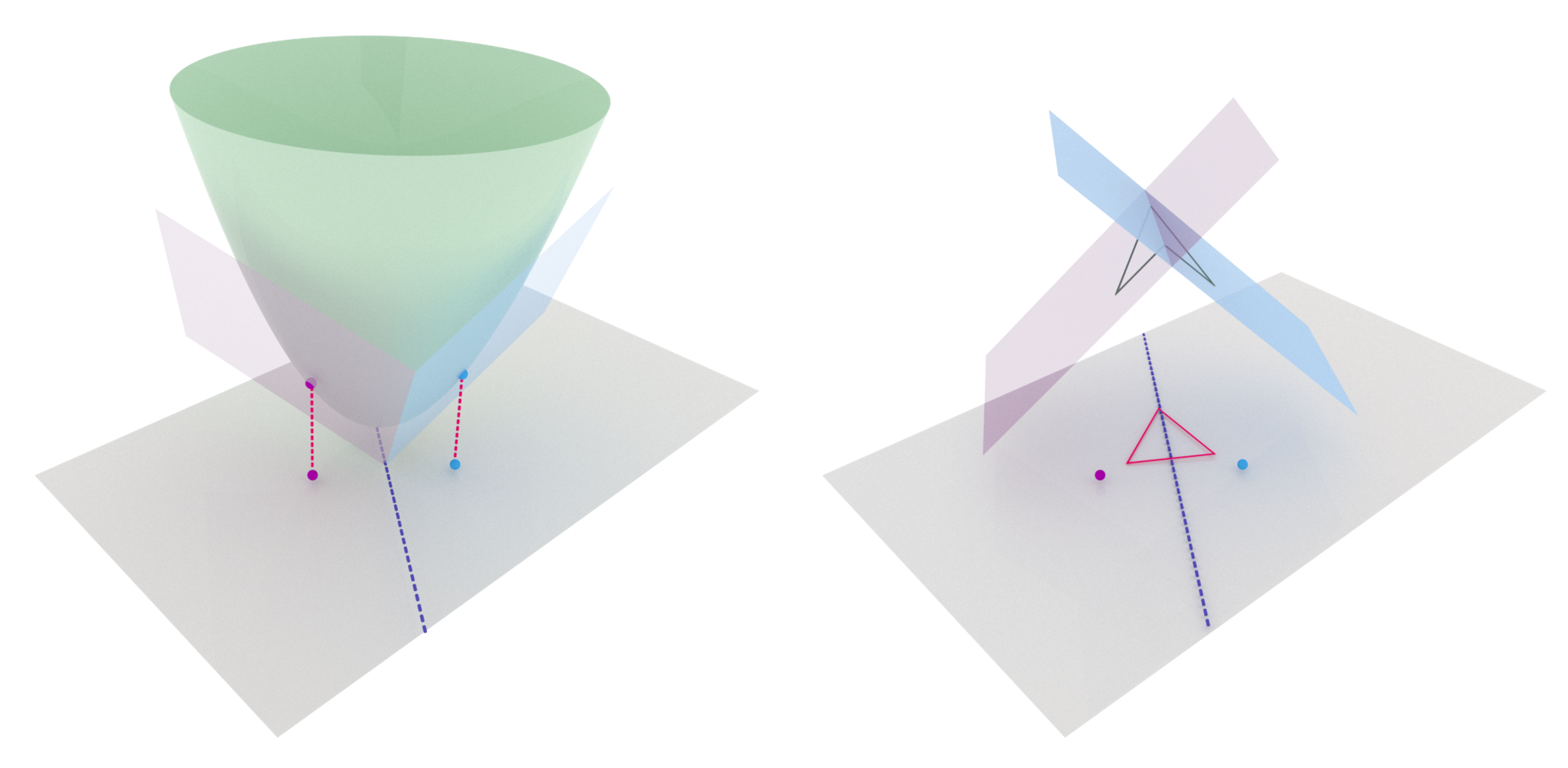}
\put(18.7,15){p}
\put(28,15){q}
\put(69,15){p}
\put(78,15){q}
\put(16,-3){\small{(a)~Parabolic lifting}}
\put(63,-3){\small{(b)~Our lifting}}
\end{overpic}
%   \includegraphics[width=\columnwidth]{figs/img4cut.png}
%   \makebox[.45\columnwidth]{(a)~Parabolic lifting}\makebox[.45\columnwidth]{(b)~Our lifting}
  %\vspace{-2mm}
  \end{center}
   \caption{Two different lifting schemes. 
   (a) The traditional lifting scheme lifts $(x,y)$ to $(x,y,x^2+y^2)$ that is located on a parabolic surface.
   It is well defined on a 2D plane but not easy to be extended to a curved surface. 
   (b)~Our lifting scheme lifts the three vertices of a triangle
   to 
   $(x_1,y_1,d^2(p,v_1))$, $(x_2,y_2,d^2(p,v_2))$, $(x_3,y_3,d^2(p,v_3))$ for the site $p$
   while lifting the three vertices
   to 
   $(x_1,y_1,d^2(q,v_1))$, $(x_2,y_2,d^2(q,v_2))$, $(x_3,y_3,d^2(q,v_3))$ for the site $q$. 
   }
   \label{fig:lift}
   \vspace{-3mm}
 \end{figure}
As Fig.~\ref{fig:lift}(a) shows, the traditional lifting scheme lifts $(x,y)$ to $(x,y,x^2+y^2)$ (a point on the parabolic surface $\mathcal{P}:z=x^2+y^2$).
Let $p,q$ be the two sites respectively.
The intersection line
   between the tangent plane $\Pi_{\mathcal{P}}^p$ at $p$'s lifted point and that at $q$'s lifted point, upon being projected onto the 2D plane, is exactly the perpendicular bisector of $p$ and $q$. The traditional lifting scheme is well defined on a 2D plane but not easy to be extended to a curved surface.

%   Generally, the upper envelope of the tangent planes at $\{p_i'\}_{i=1}^m$ (lifted points) is able to induce to the 2D Voronoi diagram w.r.t.~$\{p_i\}_{i=1}^m$.
   In contrast, our lifting scheme is much different; See Fig.~\ref{fig:lift}(b). Given a triangle $\triangle v_1v_2v_3$,
   the site $p$ contributes to a lifted plane~$\Pi_{v_1v_2v_3}^p$ passing through $(x_1,y_1,\|p-v_1\|^2)$, $(x_2,y_2,\|p-v_2\|^2)$, $(x_3,y_3,\|p-v_3\|^2).$
   Theorem~\ref{thm:our:lift} point out that our localized lifting scheme can also induce the exact 2D Voronoi diagram w.r.t.~$\{p_i\}_{i=1}^m$, independent of the choice of $\triangle v_1v_2v_3$.
   In the following, we use $z(x,y;\Pi)$ to denote the height of $\Pi$ at~$(x,y)$.

% \begin{lemma}\label{lem:parabola}
% Given a set of sites $\mathcal{P}=\{p_i\in\mathbb{R}^2\}_{i=1}^m$, the upper envelope of 
% \begin{displaymath}
% \Pi_{\mathcal{P}}^{p_i},\quad i=1,2,\cdots,m
% \end{displaymath} 
% defines the Voronoi diagram.
% \end{lemma}

\begin{lemma}\label{thm:self:dominate:v2}
For any non-degenerate triangle $\triangle v_1v_2v_3$ in the xy-plane, and any two different points $p(x_p,y_p),q(x_q,y_q)$ in the plane (may be outside $\triangle v_1v_2v_3$), we have
% \begin{displaymath}
% \Pi_{ABC}^p(p)<\Pi_{ABC}^q(p)\quad\text{and}\quad\Pi_{ABC}^q(q)<\Pi_{ABC}^p(q).
% \end{displaymath}
% Furthermore, 

\begin{displaymath}
z(x,y;\Pi_{v_1v_2v_3}^p)-z(x,y;\Pi_{v_1v_2v_3}^q) = z(x,y;\Pi_{\mathcal{P}}^q)-z(x,y;\Pi_{\mathcal{P}}^p)
\end{displaymath}
holds for any point $(x,y)$.
\end{lemma}
\begin{proof}
On the one hand,
\begin{displaymath}
z(x_1,y_1;\Pi_{v_1v_2v_3}^p)-z(x_1,y_1;\Pi_{v_1v_2v_3}^q)=\|p-v_1\|^2-\|q-v_1\|^2.
\end{displaymath}
On the other hand, 
\begin{align}
&z(x_1,y_1;\Pi_{\mathcal{P}}^q)-z(x_1,y_1;\Pi_{\mathcal{P}}^p)\nonumber\\
=&2x_q x_1 + 2y_q y_1 - x_q^2-y_q^2-(2x_p x_1 + 2y_p y_1 - x_p^2-y_p^2)\nonumber\\
=&(x_p-x_1)^2+(y_p-y_1)^2-(x_q-x_1)^2-(y_q-y_1)^2\nonumber\\
=&\|p-v_1\|^2-\|q-v_1\|^2.
\end{align}
By combining them, we have
\begin{displaymath}
z(x_1,y_1;\Pi_{v_1v_2v_3}^p)-z(x_1,y_1;\Pi_{v_1v_2v_3}^q) = z(x_1,y_1;\Pi_{\mathcal{P}}^q)-z(x_1,y_1;\Pi_{\mathcal{P}}^p),
\end{displaymath}
which also holds for $v_2$ and $v_3$.
Considering the assumption that $v_1,v_2,v_3$ are not co-linear
and the fact that both $$\left(z(x,y;\Pi_{v_1v_2v_3}^p)-z(x,y;\Pi_{v_1v_2v_3}^q)\right)$$ and $$\left(z(x,y;\Pi_{\mathcal{P}}^q)-z(x,y;\Pi_{\mathcal{P}}^p)\right)$$
linearly depend on $x$ and $y$, we finish the proof.
\end{proof}

Based on Lemma~\ref{thm:self:dominate:v2}, 
it is easy to justify the triangle-based lifting scheme (see Fig.~\ref{fig:lift}(b)) by the following theorem. 
\begin{theorem}\label{thm:our:lift}
Given $\triangle v_1v_2v_3$ in the xy-plane, as well as a set of sites $\{p_i\}_{i=1}^n$ on the plane (may be outside $\triangle v_1v_2v_3$), 
the configuration of the lower envelope of the following planes 
\begin{displaymath}
\Pi_{v_1v_2v_3}^{p_i},\quad i=1,2,\cdots,n
\end{displaymath}
defines the 2D Voronoi diagram w.r.t.~$\{p_i\}_{i=1}^n.$
Furthermore, the 2D Voronoi diagram is independent of the choice of $\triangle v_1v_2v_3$.
\end{theorem}

\begin{proof}
Suppose that we have a parabolic surface $\mathcal{P}:z=x^2+y^2$.
It's well known that the upper envelope of 
\begin{displaymath}
\Pi_{\mathcal{P}}^{p_i},\quad i=1,2,\cdots,n
\end{displaymath} 
defines the Voronoi diagram.
Based on Lemma~\ref{thm:self:dominate:v2}, the inequality
\begin{displaymath}
z(x,y;\Pi_{\mathcal{P}}^{p_i})-z(x,y;\Pi_{\mathcal{P}}^{p_j})\geq 0
\end{displaymath}
exactly implies 
\begin{displaymath}
z(x,y;\Pi_{v_1v_2v_3}^{p_i})-z(x,y;\Pi_{v_1v_2v_3}^{p_j}) \leq 0.
\end{displaymath}
In this way, we verify the conclusion.
\end{proof}

\begin{figure}
%\vspace{-3.0mm}
	\centering
\begin{overpic}
[width=.9\linewidth]{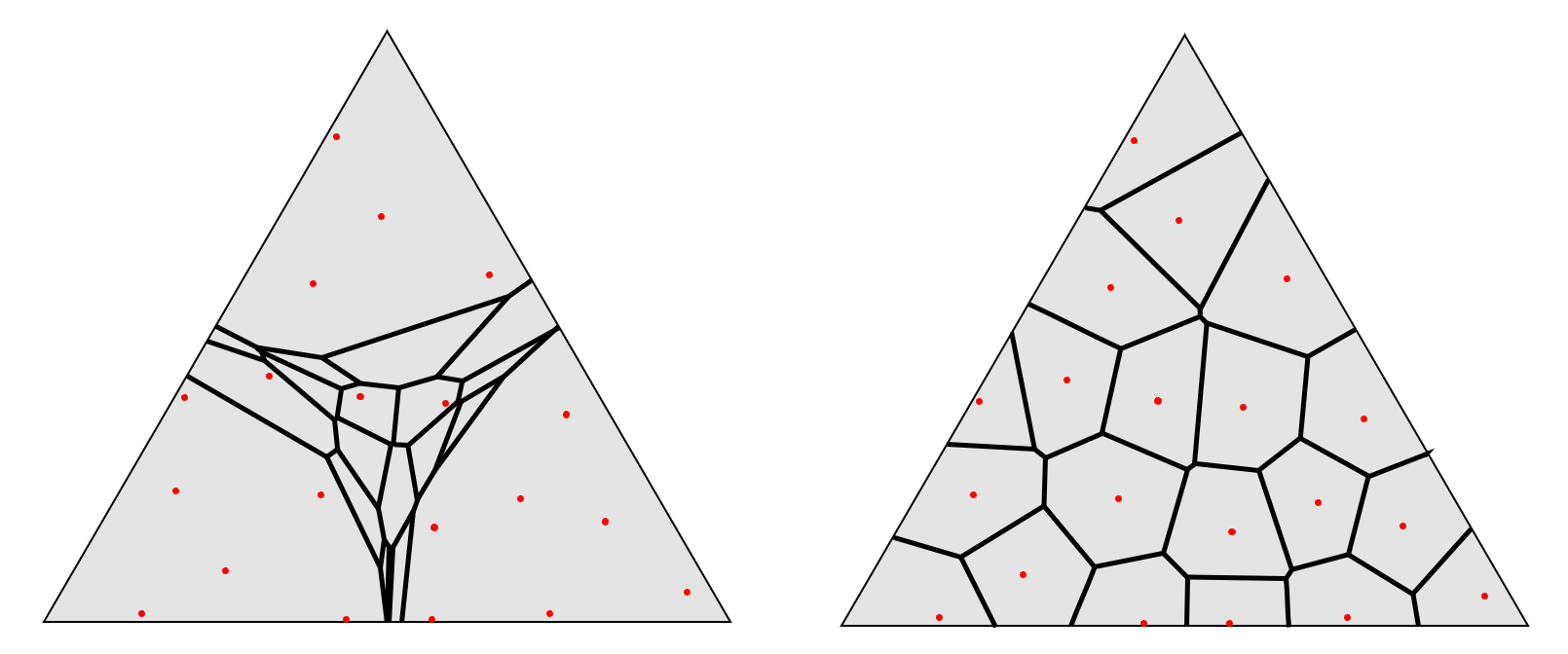}
\put(9,-3){\small{(a)~\iffalse Straight-line\fi Euclidean distance}}
\put(61,-3){\small{(b)~Squared distance}}
\end{overpic}
\caption{
(a) Algorithm~\ref{alg:mark-and-sweep} cannot report 
an exact VD when the base surface degenerates to a convex 2D region, thus lacking the property of compatibility. 
(b) By using squared distance to define half-planes,
the resulting VD is the same with the exact 2D Voronoi diagram, endowing Algorithm~\ref{alg:mark-and-sweep} with the property of compatibility and the ability of dealing with face-interior sites. 
}
\label{FIG:compatibility:squaredDistance}
 \vspace{-3mm}
\end{figure}

\begin{figure*}[h]
%\vspace{-3.0mm}
	\centering
\begin{overpic}
[width=\linewidth]{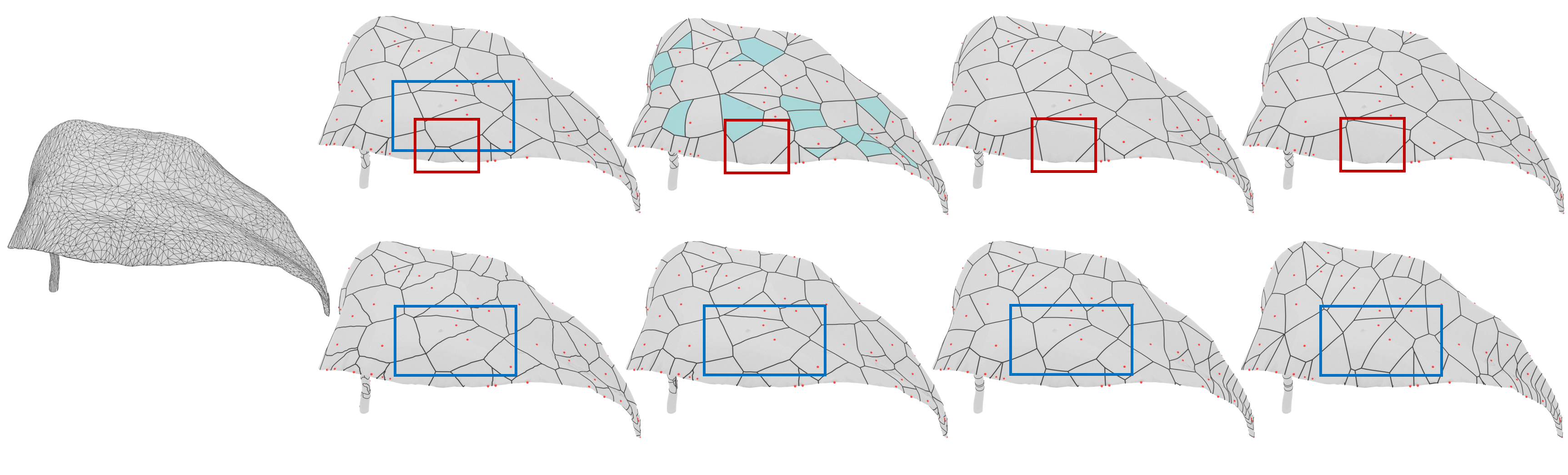}
\put(7,9){\small{(a)~Input}}
\put(23,16){\small{(b)~Ours+ExactGeodesic}}
\put(47,16){\small{(c)~RVD}}
\put(66,16){\small{(d)~LRVD}}
\put(82,16){\small{(e)~Ours+\iffalse Straight-line\fi Euclidean}}
\put(23,1){\small{(f)~Diffusion Diagram}}
\put(42,1){\small{(g)~Ours+FastMarching}}
\put(62,1){\small{(h)~Ours+Commute-time}}
\put(82,1){\small{(i)~Ours+Biharmonic}}
% \put(7,1){(f)~Diffusion Diagram}
% \put(30.5,1){(g)~Ours+FastMarching}
% \put(56,1){(h)~Ours+Commute-time}
% \put(81,1){(i)~Ours+Biharmonic}
% \put(29,21){(b)~Ours + Fast Marching}
% \put(50,21){(c)~Ours + Straight-line Distance}
% \put(75,21){(d)~Ours + Exact Geodesic Distance}
% \put(0,1){(e)~Ours + Commute-time Distance}
% \put(27,1){(f)~Ours + Biharmonic Distance}
% \put(50,1){(g)~Ours + Semi-harmonic Distance}
% \put(77,1){(h)~Ours + Triharmonic Distance}
\end{overpic}
\caption{
Taking the Leaf model as the input (a),
we visually compare the results produced by various approaches. 
We use the VTP algorithm to feed 
exact distances into our algorithm, yielding a high-quality surface VD (b). % to drive the computation of surface  
Both RVD (c) and LRVD (d) use the \iffalse straight-line\fi Euclidean distance to drive the surface partition, but RVD may produce ownerless regions (highlighted in light blue).  If we takes \iffalse straight-line\fi Euclidean distance to drive the partition, the result (e) is the same as LRVD on this model (fair triangulation). Diffusion diagram (f)
is inaccurate except that the input mesh is with a high triangulation quality. If we take the fast marching method (g) as the plugin, 
the results are similar to (b) but not accurate.
Commute-time distance (h) or biharmonic distance (i) is far from geodesic distance, thus causing a VD to be very different from (b). 
}
\label{FIG:leaf}
\vspace{-3mm}
\end{figure*}
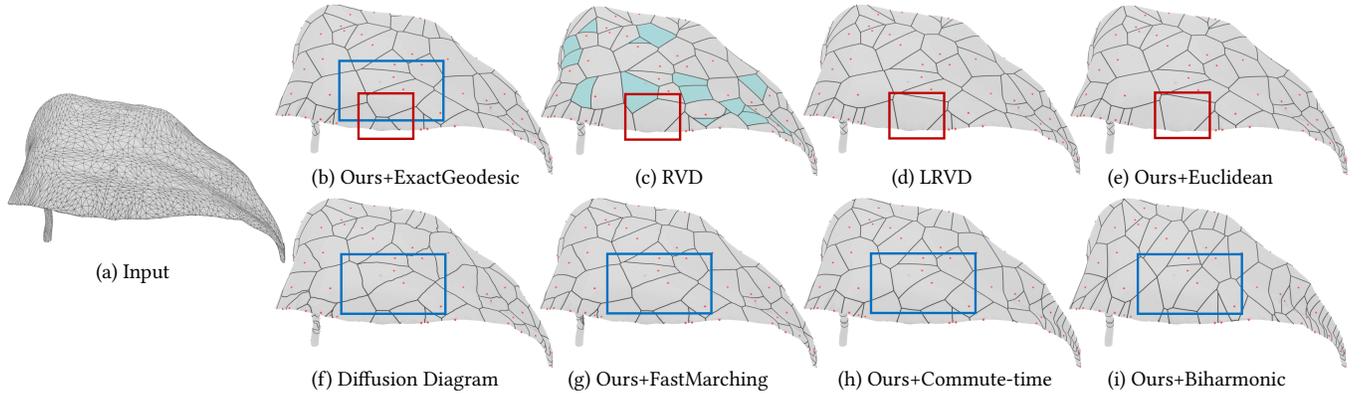

\paragraph{Remark on Theorem~\ref{thm:our:lift}}
In Section~\ref{subsec:over-propagation}, we assume that every $p$-based geodesic distance field 
changes linearly in the triangle $\triangle v_1v_2v_3$. Lemma~\ref{thm:self:dominate:v2} indicates that
it is better to assume that the squared distance, instead of the original geodesic distance or \iffalse straight-line\fi Euclidean distance, undergoes a linear 
change, i.e., the source point $p_i$ defines a plane over $\triangle v_1v_2v_3$ that passes through
\begin{displaymath}
\left(x_1,y_1,\mathbf{d}_g^2(p_i,v_1)\right),
\left(x_2,y_2,\mathbf{d}_g^2(p_i,v_2)\right),
\left(x_3,y_3,\mathbf{d}_g^2(p_i,v_3)\right).
\end{displaymath}
In this way, the property of ``compatibility'' can be guaranteed, without increasing the computational cost to Algorithm~\ref{alg:mark-and-sweep}. 
%Furthermore, it doesn't influence the sweep-and-mark propagation scheme or require additional computational cost. 
% In other words, we infer the surface-based VD by inferring how the squared-distance fields rooted at different sources compete in $\triangle v_1v_2v_3$, although 
% what we propagate Algorithm~\ref{alg:mark-and-sweep} remains the geodesic distance. 
% With this assumption, the resulting partitioning by Algorithm~\ref{alg:mark-and-sweep} is compatible with the traditional 2D Voronoi diagram,
% which is a nice feature for computing surface-based VDs.
% For example, in some mesh sampling scenarios, there are many points in one triangle. 
% It's better to ensure that the resulting diagram restricted in that triangle is the same with VD in~$\mathbb{R}^2$. 
The property of ``compatibility'' is very useful in practice.
% First, it can effectively suppress the zigzag of bisectors when one uses approximate geodesic algorithms to infer surface-based VDs;
% See Fig.~\ref{FIG:leaf}.
% Second, 
It enables Algorithm~\ref{alg:mark-and-sweep} to support face-interior sites; See
Fig.~\ref{FIG:compatibility:squaredDistance}.

We have finished decoupling the computation of surface-based VDs
from the geodesic algorithms that are used to measure the distance between two points.
By using different geodesic algorithms to measure 
the distance, we have at least three approaches for computing surface-based VDs. 
\begin{itemize}
    \item Exact geodesic Voronoi diagram (EGVD)
    that works with an exact geodesic algorithm;
    \item Approximate geodesic Voronoi diagram (AGVD)
     that works with the fast marching method;
    \item \iffalse  Straight-line\fi Euclidean distance based Voronoi diagram (EDBVD)
     that works with just \iffalse straight-line\fi Euclidean distance.
\end{itemize}

\section{Evaluation}
\label{sec:Evaluation}
\subsection{Experimental setting}
We implemented our algorithm in C++, which is independent of an off-the-shelf Voronoi/Delaunay solver,  on the  platform with a 2.8 GHz Intel Core i5-8400 CPU and Windows 10 operating system.

\subsection{Comparison with the state of the arts}
Fig.~\ref{FIG:leaf}(a)
shows a thin-sheet leaf model with 35K triangles.
By specifying 100 sites on the surface, we compare with some known algorithms including
RVD~\cite{yan2009isotropic}, LRVD~\cite{yan2014low},
and diffusion diagram~\cite{https://doi.org/10.1111/cgf.13116}. 
As our approach supports various forms of geodesic distances,
we respectively use exact geodesic distance (computed by the  VTP geodesic algorithm~\cite{qin2016fast}),
approximate geodesic distance (computed by the fast marching method~\cite{doi:10.1137/S0036144598347059}),
commute-time distance~\cite{4072747},
biharmonic distance~\cite{doi:10.1145/1805964.1805971},
% semiharmonic distance~\cite{?},
% triharmonic distance~\cite{?}\SQ{double check}\PF{deleted two graphs},
and even \iffalse  straight-line\fi Euclidean distance as the distance solvers.

\begin{figure}
%\vspace{-3.0mm}
	\centering
\begin{overpic}
% [width=\linewidth]{figs/knot6.png}
[width=\linewidth]{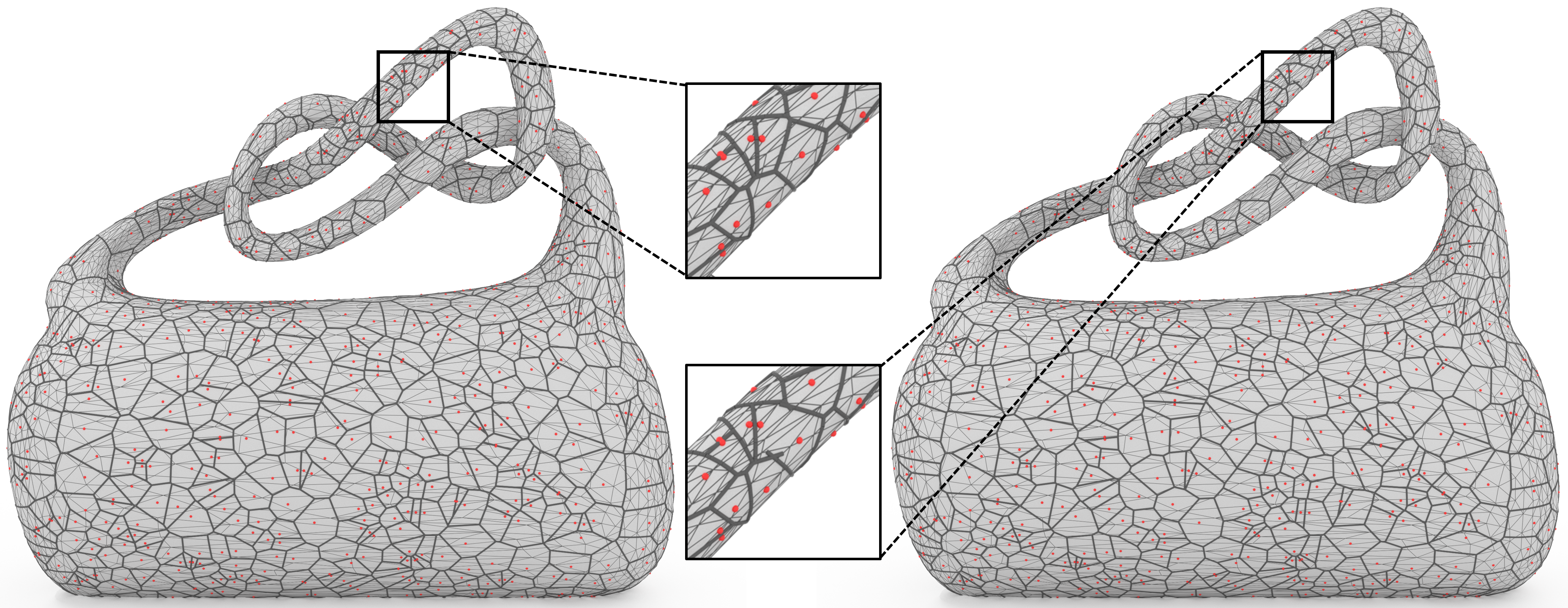}
\put(15,-3){\small{(a)~Ours}}
\put(73,-3){\small{(b)~LRVD}}
\end{overpic}
\vspace{-5mm}
\caption{
Comparison between EDBVD and LRVD on a poorly triangulated surface. 
Since 
the Dijkstra-sweep along mesh edges may fail to report reliable neighboring relationship between sites, LRVD may miss some bisectors but EDBVD cannot.}

\label{FIG:poorly:triangulated}
\vspace{-3mm}
\end{figure}

\begin{figure*}[h]
%\vspace{-3.0mm}
	\centering
\begin{overpic}
[width=\linewidth]{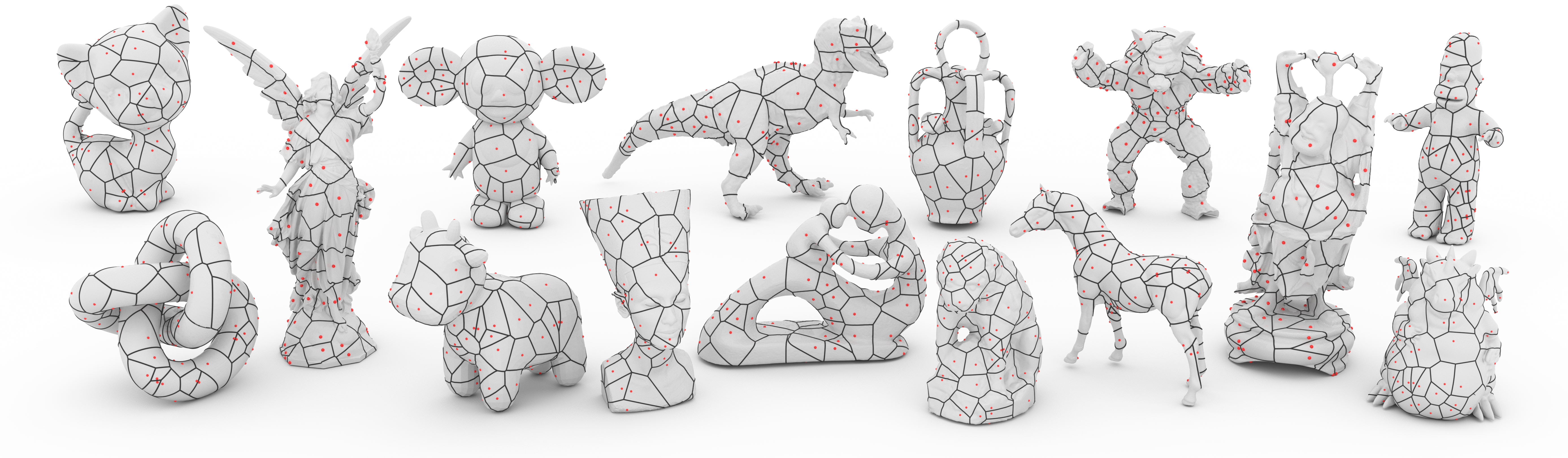}
\end{overpic}
\caption{More surface VDs produced by our EDBVD.}
\label{FIG:results}
	\vspace{-2mm}
\end{figure*}

\begin{figure*}[h]
%\vspace{-3.0mm}
	\centering
\begin{overpic}
[width=\linewidth]{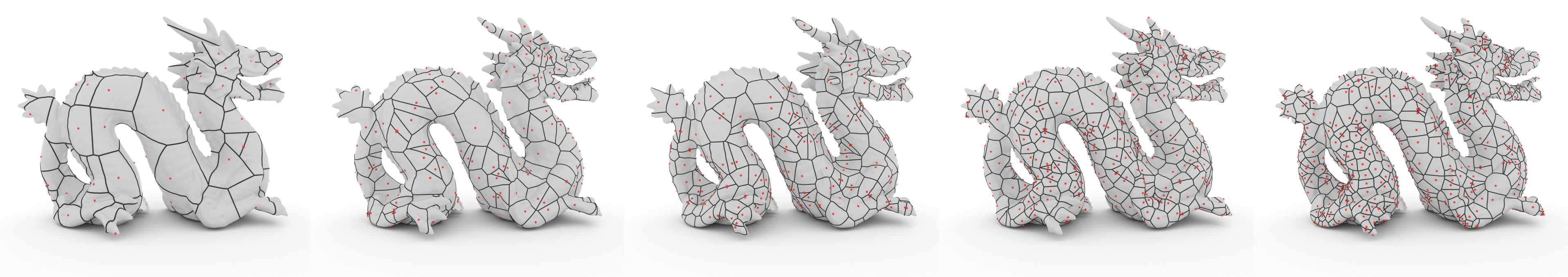}
\put(8,1){(a)~100}
\put(29,1){(b)~300}
\put(49,1){(c)~500}
\put(69,1){(d)~700}
\put(89,1){(e)~900}
\end{overpic}
\caption{Test EDBVD on the Dragon model by specifying various number of sites, where the base surface is with 50K triangles. 
}
\label{FIG:diffSources}
	\vspace{-3mm}
\end{figure*}
% \vspace{-10mm}
Among the existing algorithms,
RVD is the most commonly used one in the computer graphics community,
but the biggest disadvantage of RVD is that 
it may violate the ``one site, one region'' property, especially for those models with thin parts,
resulting in many ownerless regions;
See the highlighted regions in Fig.~\ref{FIG:leaf}(c).
Yan et al.~\shortcite{yan2014low} 
proposed to improve the traditional RVD to localized RVD (LRVD) by enforcing a Dijkstra-sweep step.
For a model with fair triangulation,
if our algorithm takes Euclidean distance to drive the partition, the result is the same with LRVD because each curved bisector in LRVD or ours is obtained by intersecting the surface with a bisector plane between a pair of sites.
But the Dijkstra-sweep along mesh edges may fail to report reliable neighboring relationship between sites on poorly triangulated surfaces; 
See~\cite{wang2020robustly} for detailed discussion.
Our EDBVD, although it looks similar to LRVD, 
sweeps the surface from triangle to triangle
while coordinating the propagation by the \iffalse straight-line\fi Euclidean distance, rather than the graph based distance reported by Dijkstra's algorithm.
Therefore, EDBVD is independent of triangulation quality; See Fig.~\ref{FIG:poorly:triangulated}.
% %%
% %%
% Unlike that RVD may produce ownerless regions,
Both LRVD and EDBVD have
the ``one site, one region'' property.
We observe that the results of LRVD and EDBVD may be different from the real GVD along the sharp edge of the geometry,
due to the significant difference between \iffalse straight-line\fi Euclidean distance and geodesic distance.

To our knowledge, the diffusion diagram algorithm~\cite{https://doi.org/10.1111/cgf.13116} is also competitive in generating surface-based VDs; See Fig.~\ref{FIG:leaf}(f).
It takes the necessity of over-propagation into consideration but lacks a good strategy on reducing the propagation range. 
Besides, it is sensitive to the triangulation quality, and
the resulting VD may be significantly different from the ground-truth when the base surface does not have a high triangle quality.
% Second, it infers the VD branching structure by repeatedly splitting a triangle until each triangle has at most two contributing sites. The splitting process is very tedious and prone to numerical errors (the triangulation quality becomes worse and worse). 

In contrast, our approach is flexible.
It supports arbitrary distance solver as the plugin. Fig.~\ref{FIG:leaf}(b,e,g,h,i) show
some surface-based VDs given by different distance solvers. 
If we take the fast marching method  as the plugin of our algorithm, 
the result is similar to Fig.~\ref{FIG:leaf}(b) but not accurate.
Commute-time distance (see Fig.~\ref{FIG:leaf}(h)) or biharmonic distance (see Fig.~\ref{FIG:leaf}(i)) is far from geodesic distance, thus leading to a VD much different from Fig.~\ref{FIG:leaf}(b).
% As Fig.~\ref{FIG:leaf}(b,c,d) show,
% our algorithm supports user-specified routines to define distances,
% leading to different surface-based VDs.
% An interesting observation is that 
% in spite of being much different in accuracy,
% both exact/approximate geodesic distances and straight-line distances 
% produce satisfactory results. 
% The rational behind lies in (1)~the sweep-and-mark algorithm paradigm effectively prevents penetrating into a thin-sheet model, (2)~the VD bisectors are mostly given by two nearby sites,
% making straight-line distance suffice to infer the position of the bisector, and (3) our algorithm is based on the strategy of finding the envelope of multiple distance fields (that is more informative) rather than simply the multi-source distance field. 
% Our SLDBVD (using straight-line distance) 
% can better solve the stumbling task with a similar timing cost; See Table~\ref{tab:Performance}.
Generally speaking, we recommend the fast marching method to drive the computation
if the sites are very sparse, while using \iffalse straight-line\fi Euclidean distance directly if the sites are very dense. 
AGVD and EDBVD are enough for the most computer graphics applications. 
Fig.~\ref{FIG:results} shows a gallery of results computed by EDBVD.
We also tested the exact geodesic distance driven GVD algorithms such as~\cite{liu2017constructing} and~\cite{xu2014polyline}.
The resulting VDs are not significantly different from AGVD, but they are hundreds of times slower than our AGVD or thousands of times slower
than our EDBVD.
The detailed ratio depends on the mesh resolution and the number of sites.

%(calling an approximate geodesic distance solver or directly using straight-line distance).

% Albut the visual difference between their VDs our VDs is inconspicuous.
% Considering that in the computer graphics community 
% the commonly used 3D models contain tens of thousands of triangles, using exact geodesic algorithms to compute surface VDs becomes impractical. 
% What's more, the propagation scheme and the computation of VDs are highly coupled for most existing GVD algorithms, greatly limiting their use.
%scalability. 

\vspace{-1.5mm}

\subsection{Run-time performance}
% \begin{figure}
% %\vspace{-3.0mm}
% 	\centering
% \begin{overpic}
% [width=\linewidth]{figs/3d1.png}
% \put(36,-2){\small{No.of Sites}}
% \put(82,3){\small{\rotatebox{45}{No.of Faces}}}
% \put(-3,18){\small{\rotatebox{90}{Time (s)}}}
% \end{overpic}
% \caption{
% Timing cost statistics by varying 
% the mesh resolution and
% the number of sites.
% }
% \label{FIG:Time3d}
% \vspace{-3mm}
% \end{figure}
Fig.~\ref{FIG:diffSources} 
shows a group of surface VD results on the Dragon model
with 50K triangles, computed by EDBVD.
By setting the number of sites to be 100, 300, 500, 700, and 900, 
the required timing costs are respectively 0.062 s, 0.094 s, 0.109 s, 0.125 s, and 0.130 s.
% which shows that the timing cost increases nearly linearly to the number of sites.

% Statistics show that
% the total timing cost basically
% linearly depends on the two dimensions. 

%$\lambda_1 O(m)+\lambda_2 O(n)$
% It's worth noting that
% among the two kinds of operations, i.e., 
% distance propagation, 
% plane cutting, 
% \SQ{too kinds of operations..}\PF{fixed} InferOverPropagatedDistancesForLRVD use 60$\%$,range from 58$\%$ to 65$\%$, Triangle_Base_Convex_Top use 33$\%$, range from 28$\%$ to 40$\%$, ratio....bottleneck

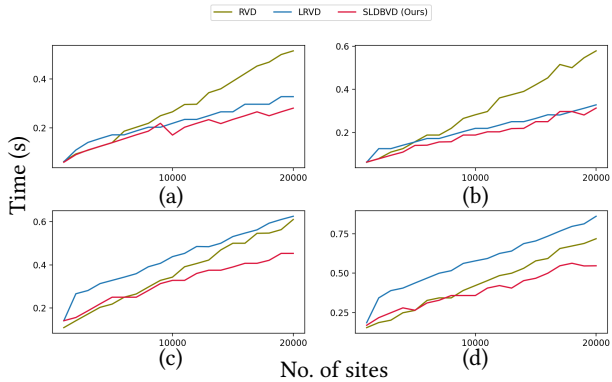
\begin{figure}
%\vspace{-3.0mm}
	\centering
\begin{overpic}
[width=.9\linewidth]{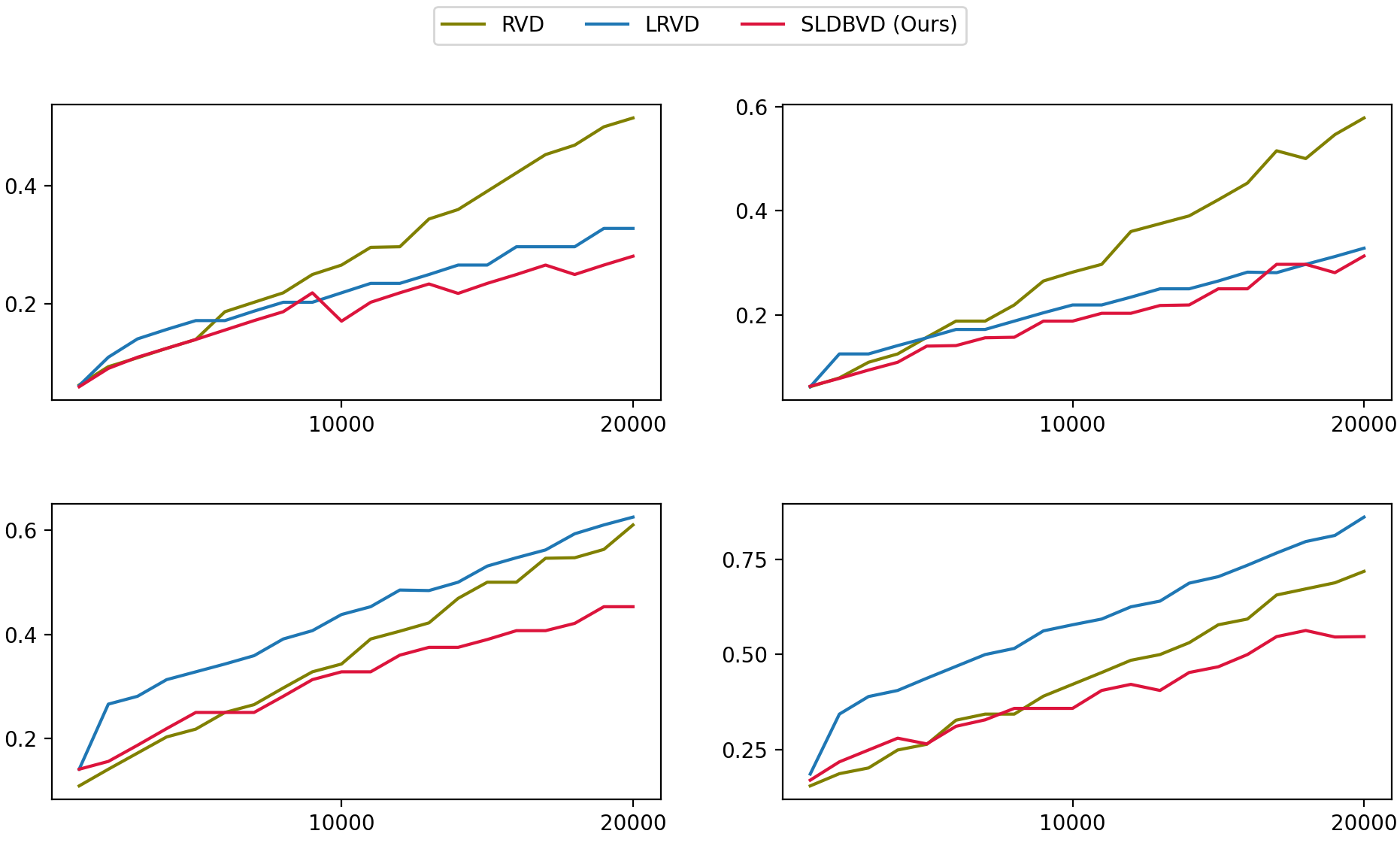}
\put(-4,25){\rotatebox{90}{Time (s)}}
\put(43,-3){No. of sites}
\put(22,27){(a)}
\put(74.5,27){(b)}
\put(22,-1){(c)}
\put(74.5,-1){(d)}
\end{overpic}
	%\vspace{-2mm}
\caption{
Run-time performance comparison among RVD, LRVD and our EDBVD.
(a) Vase: 30K faces.
(b) Dolphin: 30K faces.
(c) Bunny: 70K faces.
(d) Horse: 100K faces.
}
\label{FIG:Time}
	\vspace{-3mm}
\end{figure}

RVD, LRVD, and EDBVD are all based on \iffalse straight-line\fi Euclidean distance. 
Therefore, it is fair to compare the run-time performance among the three algorithms.
We select four models for tests, i.e., 30K-face Vase,
30K-face Dolphin, 70K-face Bunny, and 100K-face Horse. 
By specifying various number of sites on these models, 
it shows that EDBVD outperforms RVD and LRVD in run-time, especially with an increasing number of sites. 
We also record some statistics of run-time in Table~\ref{tab:Performance}, where the best time data is highlighted in bold. Both Fig.~\ref{FIG:Time} and Table~\ref{tab:Performance} validate the efficiency of EDBVD.

Next, we come to analyze the time complexity of EDBVD.
First, we assume that $n$
and $m$ are at the same order of magnitude, where $m$ is the number of sites
and $n$ is the number of faces.
We further assume that (1)~the triangulation is dense and fine, (2)~each site dominates only several triangles, and (3)~the number of sites in a triangle can be bounded by $O(1)$.
On the one hand, the time cost for conducting the propagation of each site is $O(1)$,
and thus the total timing cost spent in distance propagation
can be bounded by $O(m)$.
Additionally, each triangle receives a limited number of distance triples, generally between 1 to 3, if no serious coincidence occurs.
The total timing cost spent in half-plane cutting can be bounded by $O(n)$. 
Therefore, the average time complexity of our EDBVD algorithm 
is $O(m+n)$.
If $m\ll n$, then the total timing cost spent in distance propagation 
may amount to $O(n\log n)$. In this case, the average time complexity of EDBVD becomes $O(n\log n)$.
If $m\gg n$, instead, then the propagation cost is $O(m)$ (in the average case) and the cost of half-plane cutting amounts to $O(n\times (\frac{m}{n})^2)$ (see Section~\ref{subsec:over-propagation}), making the total time complexity $O(\frac{m^2}{n})$, which indicates that a highly efficient half-plane cutting solver is necessary when $m$ is large. 

Finally, we must point out that if a different distance solver is used, the time complexity must be different. We come to analyze the average time complexity under the assumption of fair triangulation and uniform distribution of sites.
Suppose that we are using the fast marching method, running in time~$O(n\log n)$, as the distance solver. If $m\ll n$, the average time complexity climbs to $O(n\log n)$.
For $m=O(n)$ and $m\gg n$, the time complexity remains unchanged,
i.e., $O(m+n)$ for  $m=O(n)$ and $O(\frac{m^2}{n})$ for $m\gg n$.

\begin{table}
\caption{Performance statistics of RVD, LRVD and our EDBVD (in seconds), where the best run-time performance is highlighted in bold.} 
\label{tab:Performance}
\vspace{-2mm}
\resizebox{\linewidth}{!}
{
\begin{tabular}{l|c|l|l|l|l|l|l} 
\hline
\multicolumn{1}{c|}{\multirow{2}{*}{Model}} & \multirow{2}{*}{Faces}                     & \multicolumn{1}{c|}{\multirow{2}{*}{Method}} & \multicolumn{5}{c}{Number of sampling points}                                                                                                                                               \\ 
\cline{4-8}
\multicolumn{1}{c|}{}                       &                                            & \multicolumn{1}{c|}{}                        & \multicolumn{1}{c|}{4K}             & \multicolumn{1}{c|}{8K}             & \multicolumn{1}{c|}{12K}            & \multicolumn{1}{c|}{16K}            & \multicolumn{1}{c}{20K}             \\ 
\midrule
\multicolumn{1}{c|}{\multirow{3}{*}{Vase}}  & \multirow{3}{*}{30K}                       & RVD                                          & 0.126                               & 0.219                               & 0.297                               & 0.422                               & 0.515                               \\ 
\cline{3-8}
\multicolumn{1}{c|}{}                       &                                            & LRVD                                         & 0.157                               & 0.203                               & 0.235                               & 0.297                               & \multicolumn{1}{c}{0.328}           \\ 
\cline{3-8}
\multicolumn{1}{c|}{}                       &                                            & EDBVD(Ours)                                  & \textbf{0.125}                      & \multicolumn{1}{c|}{\textbf{0.187}} & \textbf{0.219}                      & \multicolumn{1}{c|}{\textbf{0.250}} & \textbf{0.281}                      \\ 
\hline
\multicolumn{1}{c|}{\multirow{3}{*}{Bunny}} & \multirow{3}{*}{70K}                       & RVD                                          & \multicolumn{1}{c|}{\textbf{0.203}} & \multicolumn{1}{c|}{0.297}          & \multicolumn{1}{c|}{0.406}          & \multicolumn{1}{c|}{0.501}          & \multicolumn{1}{c}{0.610}           \\ 
\cline{3-8}
\multicolumn{1}{c|}{}                       &                                            & LRVD                                         & \multicolumn{1}{c|}{0.313}          & \multicolumn{1}{c|}{0.391}          & \multicolumn{1}{c|}{0.485}          & \multicolumn{1}{c|}{0.547}          & \multicolumn{1}{c}{0.625}           \\ 
\cline{3-8}
\multicolumn{1}{c|}{}                       &                                            & EDBVD(Ours)                                  & \multicolumn{1}{c|}{0.219}          & \multicolumn{1}{c|}{\textbf{0.281}} & \multicolumn{1}{c|}{\textbf{0.361}} & \multicolumn{1}{c|}{\textbf{0.407}} & \multicolumn{1}{c}{\textbf{0.453}}  \\ 
\hline
\multicolumn{1}{c|}{\multirow{3}{*}{Horse}}                      & \multirow{3}{*}{100K}                      & RVD                                          & \textbf{0.253}                      & \textbf{0.344}                      & 0.485                               & 0.593                               & 0.718                               \\ 
\cline{3-8}
                                            &                                            & LRVD                                         & 0.406                               & 0.516                               & 0.625                               & 0.734                               & \multicolumn{1}{c}{0.863}           \\ 
\cline{3-8}
                                            &                                            & EDBVD(Ours)                                  & 0.281                               & 0.359                               & \textbf{0.422}                      & \textbf{0.547}                      & \textbf{0.550}                      \\ 
\hline
\multicolumn{1}{c|}{\multirow{3}{*}{Dolphin}}                    & \multicolumn{1}{l|}{\multirow{3}{*}{180K}} & RVD                                          & \textbf{0.341}                      & \textbf{0.465}                      & 0.654                               & 0.801                               & 0.942                               \\ 
\cline{3-8}
                                            & \multicolumn{1}{l|}{}                      & LRVD                                         & 0.619                               & 0.781                               & 0.957                               & 1.070                               & 1.250                               \\ 
\cline{3-8}
                                            & \multicolumn{1}{l|}{}                      & EDBVD(Ours)                                  & 0.401                               & 0.501                               & \textbf{0.583}                      & \textbf{0.660}                      & \textbf{0.742}                      \\ 
\hline
\multicolumn{1}{c|}{\multirow{3}{*}{Kitten}}                     & \multicolumn{1}{l|}{\multirow{3}{*}{270K}} & RVD                                          & \textbf{0.593}                      & 0.756                               & 0.945                               & 1.323                               & 1.667                               \\ 
\cline{3-8}
                                            & \multicolumn{1}{l|}{}                      & LRVD                                         & 0.672                               & 0.922                               & 1.141                               & 1.391                               & 1.656                               \\ 
\cline{3-8}
                                            & \multicolumn{1}{l|}{}                      & EDBVD(Ours)                                  & 0.609                               & \textbf{0.641}                      & \textbf{0.781}                      & \textbf{0.843}                      & \textbf{0.921}                      \\
\hline
\end{tabular}
}
\vspace{-3mm}
\end{table}
\iffalse
\begin{table}[]
\caption{vs other method}
\label{tab:vsOtherMethod}
\vspace{-2mm}
\resizebox{\linewidth}{!}
{
\begin{tabular}{|c|l|l|l|l|}
\hline
\multicolumn{1}{|l|}{} & speed                 & thin-plate             & equal 2D voronoi       & deal with other graph  \\ \hline
RVD                    & \multicolumn{1}{c|}{} &                        &                        &                        \\ \hline
LRVD                   &                       &                        & \multicolumn{1}{c|}{×} & \multicolumn{1}{c|}{×} \\ \hline
EDBVD(Ours)           &                       & \multicolumn{1}{c|}{√} & \multicolumn{1}{c|}{√} & \multicolumn{1}{c|}{√} \\ \hline
\end{tabular}
}
\end{table}
\fi

% \begin{figure}
% %\vspace{-3.0mm}
% 	\centering
% \begin{overpic}
% [width=.9\linewidth]{figs/differentNumFaces.png}
% \end{overpic}
% 	\vspace{-2mm}
% \caption{
% Time performance (dragon Model,1000 sites)
% }
% \label{FIG:Time}
% %	\vspace{-4mm}
% \end{figure}

\subsection{Extendability}
Our algorithm is able to deal with various variants of surface-based partition problems including 
breakline constraints, face-interior point sites, curve-type sites, density-equipped surfaces, and even surface restricted power diagram. 
% In this section, we talk about the uses of SLDBVD in 
% varying versions of surface-based partitioning problems
% that seem not easy to be handled by the existing VD solvers.% cannot handle.

\begin{figure}
%\vspace{-3.0mm}
	\centering
\begin{overpic}
[width=.9\linewidth]{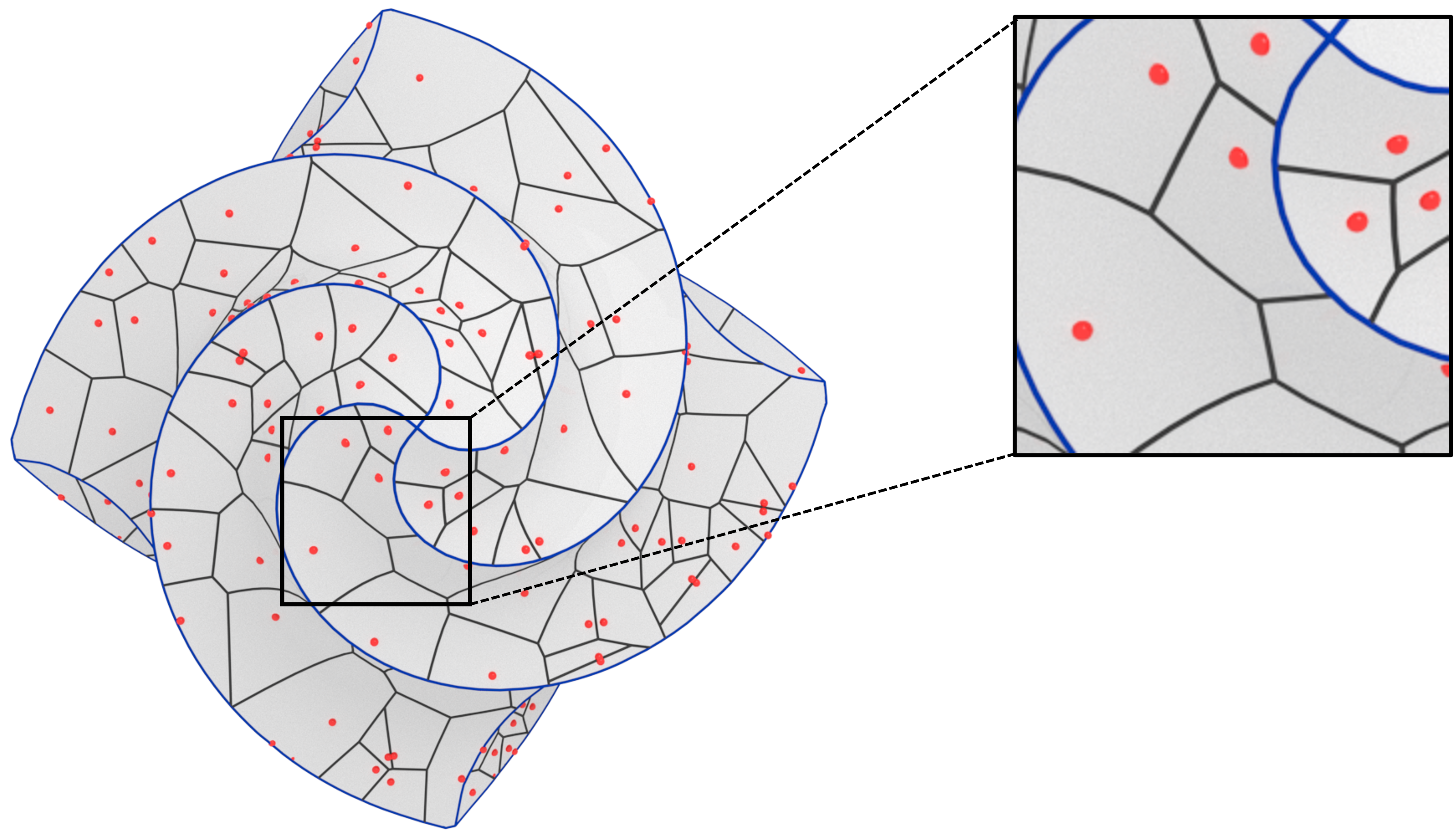}
\end{overpic}
\caption{Surface VDs with feature-line constraints. 
}
\label{FIG:constrained}
	\vspace{-3mm}
\end{figure}
\vspace{-3mm}
\paragraph{Surface VDs with breakline constraints}
There are many scenarios where one inputs a set of curves to constrain the computation of VDs~\cite{tournois20102d}. 
For example, there is a requirement in mesh generation
that VD cells are not allowed to extend across feature lines.
In fact, our algorithm can solve the stumbling problem easily. 
%naturally supports user-specified constraining curves.
Imagining that there is a linear distance field that defines a half-plane~$\pi$ over the triangle $f$, and $f$ contains a line-segment breakline. 
We only need to use the vertical plane passing through the line-segment breakline as a barrier, and prevent $\pi$ crossing the vertical plane. 
Fig.~\ref{FIG:constrained} shows
a model with 16K faces
and there are 300 sites on the surface. 
The feature lines are colored in blue.  
By taking these feature lines as breaklines,
our algorithm is able to preserve the feature lines 
in computing the surface VD. %preserving meshing.
For this example, the total computational time is just 62ms. 
% It can be further pointed 
% out that our algorithm can deal with the scenario that users draw some breaklines to prevent any Voronoi cell across the breaklines, which is useful in interactive partitioning of 3D models~\cite{jadoon2018interactive}.

\begin{figure}
%\vspace{-3.0mm}
	\centering
\begin{overpic}
[width=.9\linewidth]{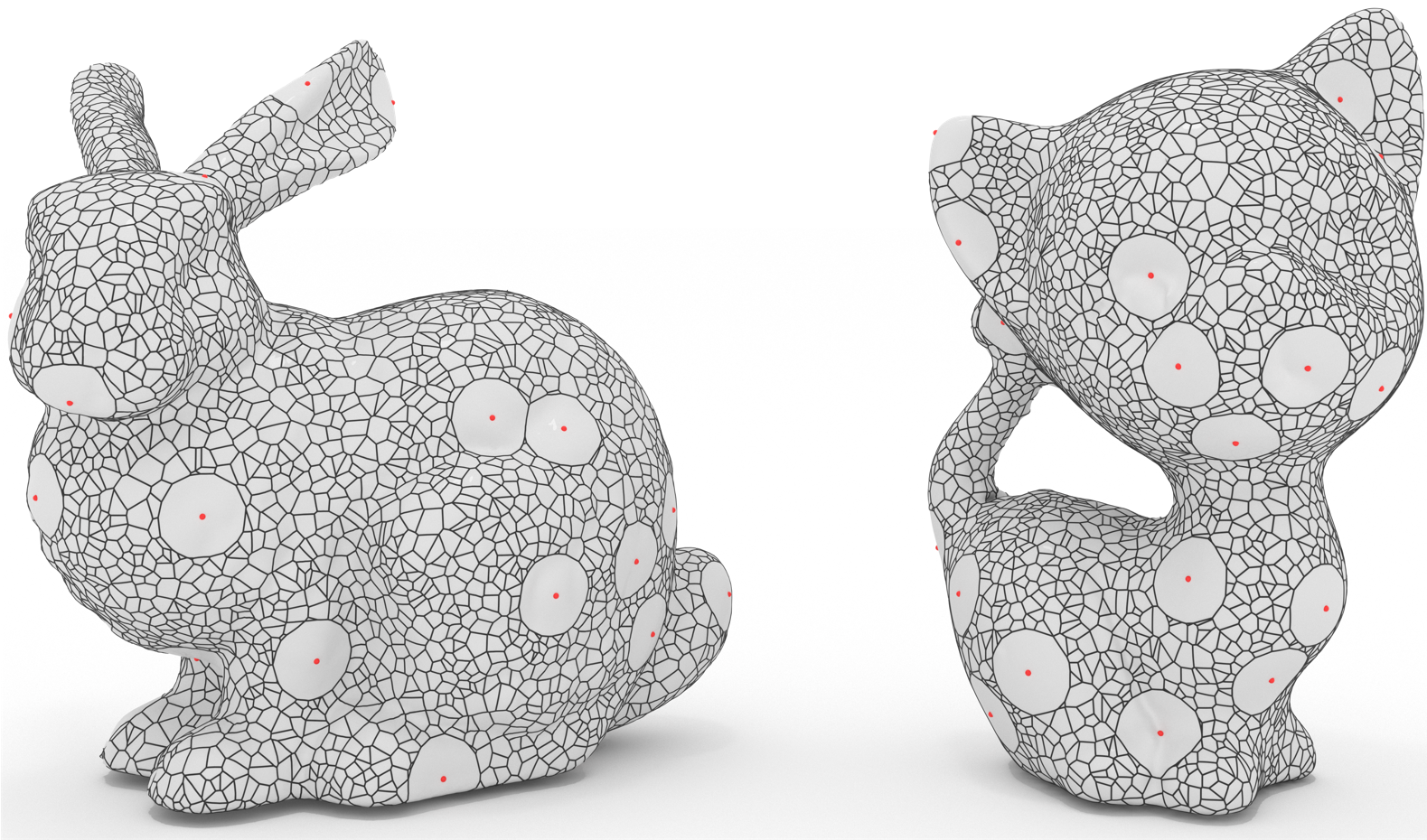}
\end{overpic}
\caption{Surface restricted power diagram.
}
\label{FIG:restricted:power}
	\vspace{-3mm}
\end{figure}
\paragraph{Restricted power diagram}
Our algorithm can be used to compute surface restricted power diagrams; See Fig.~\ref{FIG:restricted:power}.
For each vertex~$v$, 
we keep the squared distance in our implementation, i.e.,
$\|v-p\|^2,$ where $p$ is a site.
If $p$ has a weight of~$w$, then according to the definition of power diagram,
the weighted distance to~$v$ given by $p$ becomes
$\|v-p\|^2-w.$
The linear distance field in the triangle~$\triangle v_1v_2v_3$ 
is then given by 
$$
\|v_1-p\|^2-w,\|v_2-p\|^2-w,\|v_3-p\|^2-w.
$$
In this way, the bisectors of restricted power diagram can be still computed by intersecting half-planes. 
For the two examples shown in Fig.~\ref{FIG:restricted:power},
i.e., 70K-face Bunny with 6K sites and 270K-face Kitten with 5K sites,
the total computational costs are 0.234 s and 0.633 s, respectively.

\begin{figure}
%\vspace{-3.0mm}
	\centering
\begin{overpic}
[width=.9\linewidth]{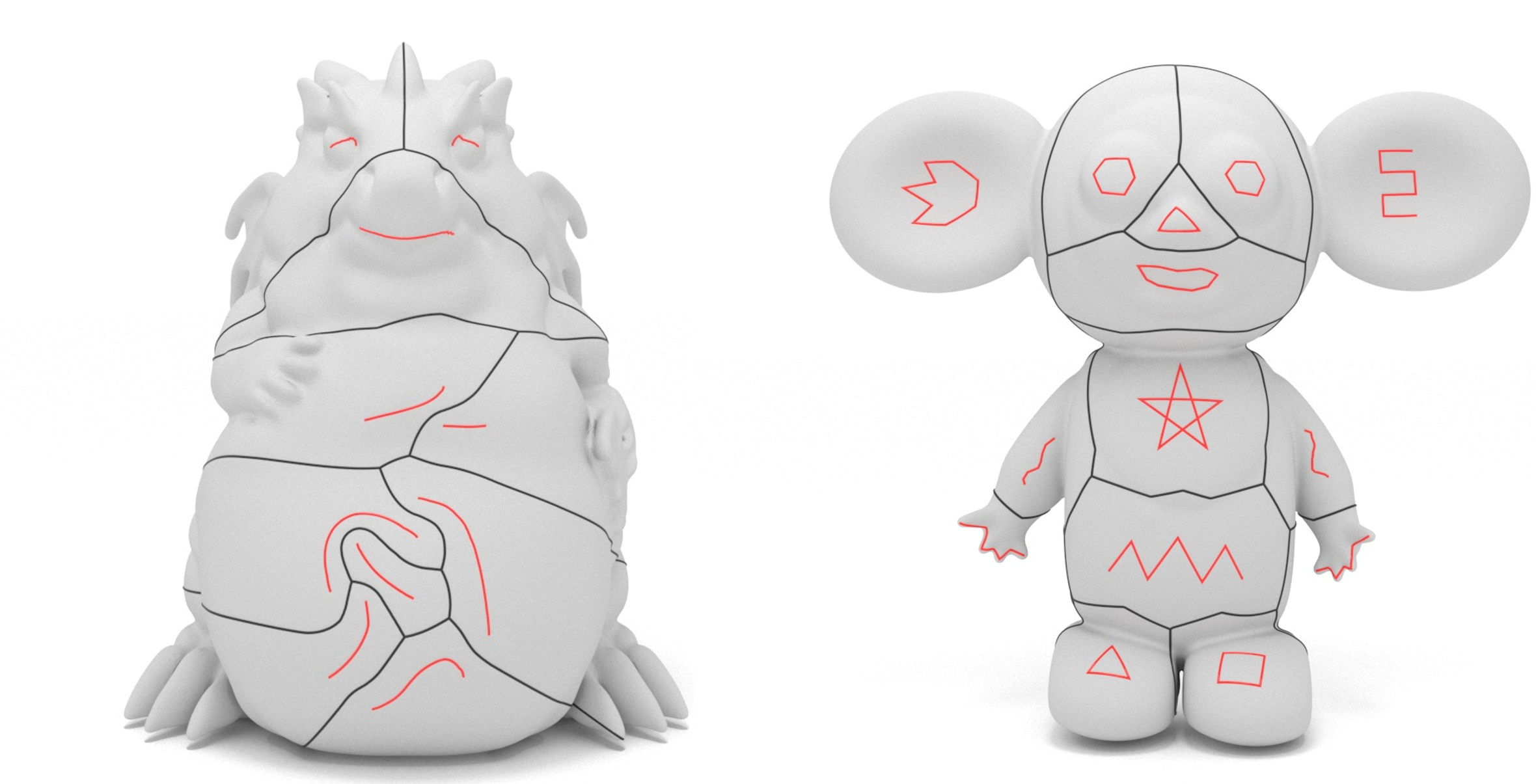}
\end{overpic}
\caption{Our algorithm supports curve-type sites.
}
\label{FIG:curve:source}
	\vspace{-3mm}
\end{figure}

\paragraph{Curve-type sites}
The VD bisectors
w.r.t. curve-type sites
are more curved than the situation of point-type sites,
making the computation of VDs very challenging.
% Computing VDs w.r.t. polyline generators
% is a challenging problem since the VD structure may contain line segments, hyperbolic segments and parabolic segments.
% As pointed out in~\cite{xu2014polyline},
% computing surface-based VDs
% is much more complicated than conventional 2D Euclidean Voronoi diagrams  as well as point-source based GVDs, since a typical bisector contains line segments, hyperbolic segments and parabolic segments.
Xu et al.~\shortcite{xu2014polyline} proposed a local Voronoi diagram (LVD) to deal with curve-type generators,
which is mainly based on the technique of additively weighted Voronoi diagram and requires a high computational cost.
For a point-type site~$p$,
we keep the squared distance, at a mesh vertex~$v$, in our implementation, i.e.,
$\|v-p\|^2.$
If the point-type sites become curve-type sites, 
it is enough to replace the point-to-point distance
with the point-to-curve distance.
The simple trick produces a good approximation to the VD of the curve-type sites; See 
Fig.~\ref{FIG:curve:source}. 
% users freely draw 10 curves. 
% Our algorithm does not need any change for dealing with curve (or polyline) sites
% except that for any mesh vertex~$v$, one need to compute the straight-line distance between a curve and $v$, rather than the distance between two points.
% At the same time, 
We must point out that RVD includes a step of calling an off-the-shelf Delauany triangulation solver, 
and thus cannot deal with curve-type sites.

\begin{figure}
%\vspace{-3.0mm}
	\centering
\begin{overpic}
[width=\linewidth]{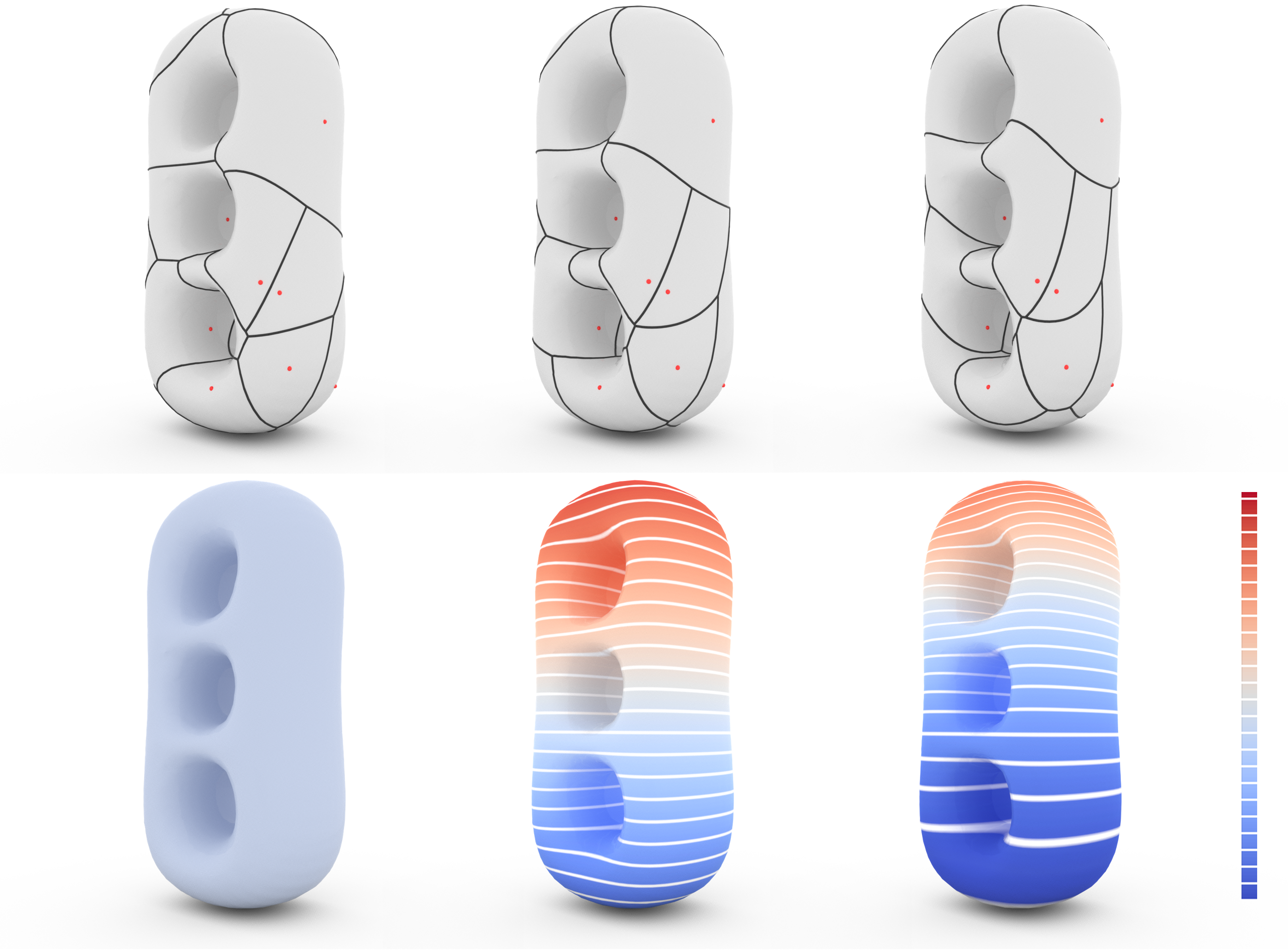}
\put(4,-3){(a)~Constant density}
\put(37,-3){(b)~Linear density}
\put(66,-3){(c)~Non-linear density}
\put(91.2,4){0.0}
\put(91.2,33){1.0}
\end{overpic}
\caption{Surface VDs on the genus-3 Torus model with different density settings.
%with 30K faces.
Our algorithm supports the input surface equipped with a non-uniform density field.
% Non-uniform density field cannot lead to any additional computational cost.
}
\label{FIG:Field}
	\vspace{-3mm}
\end{figure}

\begin{figure}
%\vspace{-3.0mm}
	\centering
\begin{overpic}
[width=\linewidth]{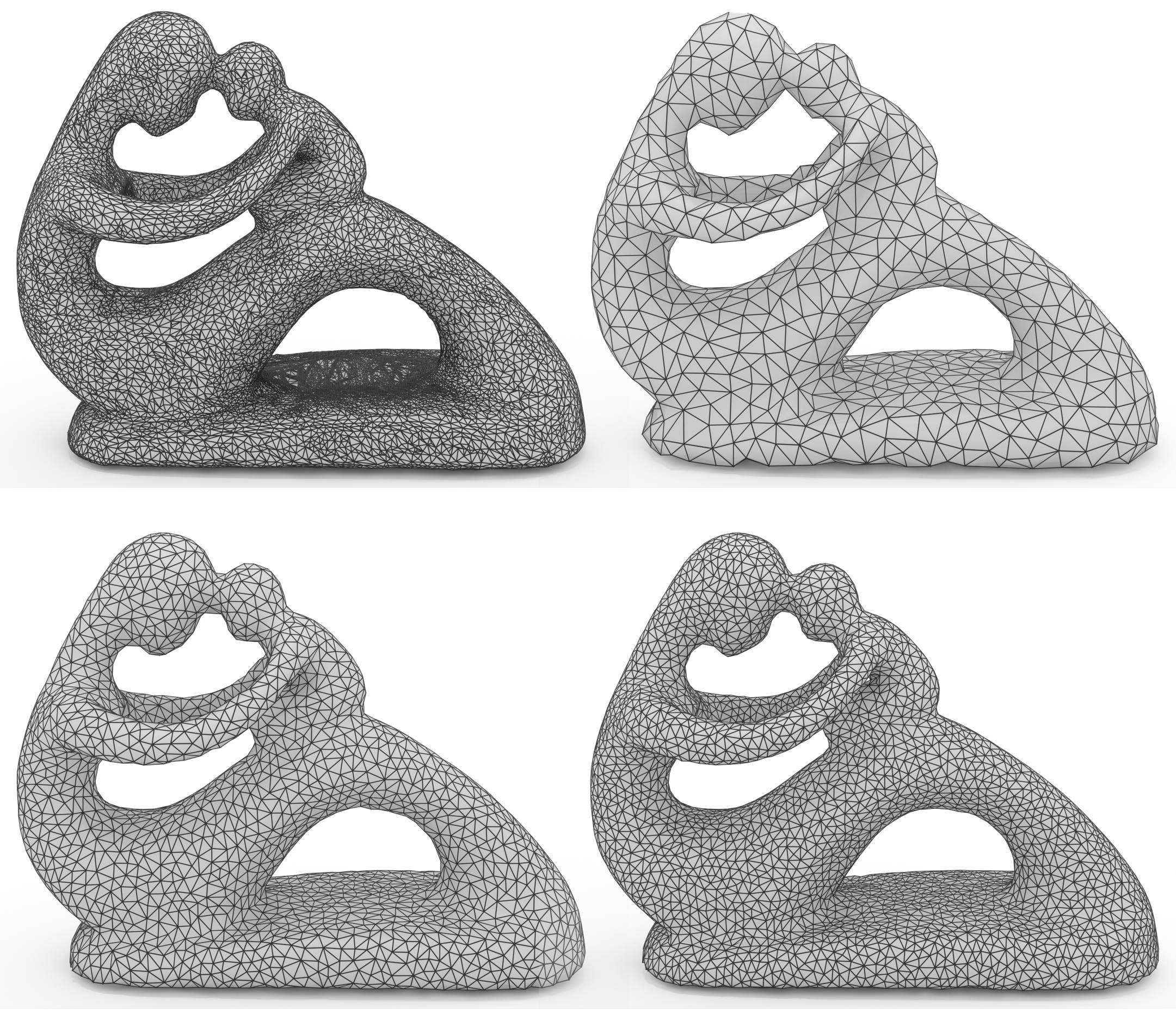}
\put(12,42){(a)~Original mesh}
\put(59,42){(b)~Remesh with 1K vertices}
\put(12,-3){(c)~Remesh with 4K vertices}
\put(59,-3){(d)~Remesh with 6K vertices}
\end{overpic}
\caption{We validate the numerical robustness of EDBVD in remeshing. 
By placing 1K, 1.1K, 1.2K, $\cdots$, 6K blue-noise sites, respectively,
our implementation does not encounter a failure case.
}
\label{FIG:Remesh}
	\vspace{-3mm}
\end{figure}

\paragraph{VD on a surface with density field}
In many digital geometry processing applications, 
the surface is equipped with a density field.
In the default setting, the density field is identity.
But the density field may also be the local-feature-size function,
shape diameter function,
mean curvature, or a kind of important field.
A basic requirement is 
to distribute sample points to adapt to 
the given density field. 
To our best knowledge,
there is no density-aware Voronoi/Delaunay solver available yet,
limiting the use of RVD in  handling the density equipped surfaces.
% Since RVD depends on the existing Delaunay solver to report the neighboring relationship, it seems not easy to handle the density equipped surfaces.

Under the assumption that the density function can be extended from the surface to 3D, i.e., $\rho=\rho(x,y,z)$,
our EDBVD can easily address the challenge. 
Let $v$ be a mesh vertex and $p$ be one of the source points.
It suffices to compute the distance as
$$
d_\rho(v,p)=d(v,p)\int_0^1 \rho(t)\text{d}t,
$$
where $d(v,p)$ is the \iffalse  straight-line\fi Euclidean distance between $v$ and $p$,
$\rho(t)$ is the density value
at the point~$(1-t)v+tp$ (a point on the line segment $\overline{vp}$), and $d_\rho(v,p)$
is the density-equipped distance.
Based on this observation,
we can use the squared density-equipped distance to define half-planes, yielding density-equipped VDs.
In Fig.~\ref{FIG:Field}, we show a genus-3 Torus model with 30K faces,
as well as the VDs for three different density settings.
It can be seen that the bisectors tend to move toward the regions with larger density. 
Note that it costs only 35 ms to compute a density-equipped VD on this model.

\paragraph{Validation on numerical robustness}
To validate our algorithm in numerical robustness, we run EDBVD
to accomplish the remeshing task by placing 
a various number of sites, which only needs to add a step of extracting the dual of the VD. 
We take the Fertility model with 36K triangles (see Figure~\ref{FIG:Remesh}(a))
as the base surface and place 1K, 1.1K, 1.2K, $\cdots$, 6K sites following the blue-noise distribution, respectively. 
No failure case occurs, which validates the numerical robustness of our algorithm. 
Figure~\ref{FIG:Remesh}(b,c,d) show
the remeshed results for 1K, 4K, and 6K sites, respectively. 
The timing costs \iffalse(sampling, running EDBVD, and extracting the dual)\fi for the three results are 0.082 s, 0.152 s, and 0.189 s, respectively.

\section{Conclusion, Limitations and Further Work}
In this paper, we propose a powerful algorithm, named SurfaceVoronoi, for computing surface-based Voronoi diagrams. 
The distinguished features of SurfaceVoronoi are two-fold:
(1)~using the over-propagation mechanism to 
keep one or more necessary distance triples, for each triangle, that may help determine the Voronoi bisectors,
and (2)~using the squared distance to define the linear change of distance in a triangle, which is a provably meaningful extension from the traditional 2D Voronoi diagram to curved surfaces.
SurfaceVoronoi has several nice features. 
First, it can work with an arbitrary geodesic algorithm. Especially, our algorithm, when driven by \iffalse straight-line\fi Euclidean distance,
is able to deal with thin-sheet models and runs faster than RVD and LRVD. 
Second, it does not need an existing Voronoi/Delaunay solver
and is flexible enough to deal with various VD problems including 
breakline constraints, face-interior point sites, curve-type sites, density-equipped surfaces, and even surface restricted power diagram. 

Our algorithm, still needs to be improved. 
First, the plane cutting operation is un-optimized and can be accelerated.
Second, if a time-consuming distance solver is used (e.g., an exact geodesic algorithm that is empirically $O(n\sqrt{n})$), the overall time cost is huge especially when the base surface contains a large number of triangles but just a few sites on the surface. 
Last but not least, numerical issues may occur when very long-and-thin triangles exist.
In the future, we will explore the possibility of 
 speeding up the computation and 
extending SurfaceVoronoi to broken meshes.
\begin{acks}
The authors would like to thank the anonymous reviewers for their valuable comments and suggestions. This work is supported by National Key R\&D Program of China (2021YFB1715900), National Natural Science Foundation of China (62272277, 62002190, 62172415, 62072284) and NSF of Shandong Province (ZR2020MF036).
\end{acks}

\bibliographystyle{ACM-Reference-Format}
\bibliography{main}
\end{document}